\documentclass[a4paper,fleqn,usenatbib]{mnras}

\usepackage{newtxtext,newtxmath}

\usepackage[T1]{fontenc}
\usepackage{ae,aecompl}

\usepackage{subfig}
\usepackage{graphicx}	
\usepackage{amsmath}	

\def\abs#1{\vert#1\vert}

\def\dch#1{{#1}}
\def\alv#1{{#1}}
\def\sr#1{{#1}}

\def\mumax{\eta_{max}}
\def\gtorder{\mathrel{\raise.3ex\hbox{$>$}\mkern-14mu
             \lower0.6ex\hbox{$\sim$}}}
\def\ltorder{\mathrel{\raise.3ex\hbox{$<$}\mkern-14mu
             \lower0.6ex\hbox{$\sim$}}}
\def\vphi{{v_{\phi}}}
\def\vphibar{\bar{v}_{\phi}}

\def\rd{{\rm d}}
\def\re{{\rm e}}
\def\ri{{\rm i}}
\def\OILR{\Omega_{\mathrm{ILR}}}
\def\tbn{(n_1,n_2)}
\newcommand{\comment}[1]
{{\color{blue} #1}} 

\title[Stability of Rotating Anisotropic Models]{The Kinematic Richness of Star Clusters - II. Stability of Spherical Anisotropic Models with Rotation}

\author[Breen, Rozier, Heggie \& Varri.]{
Philip G. Breen,$^{1}$\thanks{E-mail: philipbreen@gmail.com
}
Simon Rozier,$^{2}$
Douglas C. Heggie,$^{1}$
Anna Lisa Varri$^{3,1}$
\\
$^{1}$School of Mathematics and Maxwell Institute for Mathematical Sciences, University of Edinburgh, Kings Buildings, Edinburgh EH9 3FD\\
$^{2}$Universit\'e de Strasbourg, CNRS UMR 7550, Observatoire astronomique de Strasbourg, 11 rue de l'Universit\'e, 67000 Strasbourg, France\\
$^{3}$Institute for Astronomy, University of Edinburgh, Royal Observatory, Blackford Hill, Edinburgh EH9 3HJ, UK
}

\date{Accepted 2021 February 4. Received 2021 February 2; in original form 2020 June 29}

\pubyear{2020}

\begin{document}
\label{firstpage}
\pagerange{\pageref{firstpage}--\pageref{lastpage}}
\maketitle

\begin{abstract}
We study the bar instability in collisionless, rotating, anisotropic, stellar systems, using  $N$-body simulations and also the matrix technique for calculation of modes with the perturbed collisionless Boltzmann equation.  These methods are applied to spherical systems with an initial Plummer density distribution, but modified kinematically in two ways: the velocity distribution is tangentially anisotropic, using results of Dejonghe, and the system is set in rotation by reversing the velocities of a fraction of stars in various regions of phase space, \dch{\`a la Lynden-Bell}.  The aim of the $N$-body simulations is first to survey the parameter space, and, using those results, to identify regions of phase space (by radius and orbital inclination) which have the most important influence on the bar instability.  The matrix method is then used to identify the resonant interactions in the system which have the greatest effect on the growth rate of a bar.  Complementary series of $N$-body simulations examine these processes in relation to the evolving frequency distribution and the pattern speed.  Finally, the results are synthesised with an existing theoretical framework, and used to consider the old question of constructing a stability criterion.  
\end{abstract}




\section{Introduction}\label{sec:introduction}

The analysis of the stability properties of self-gravitating rotating equilibria is a classical problem in fluid and stellar dynamics (\citealp{Cha69}), starting with the study of the ellipsoidal figures of equilibrium. The sequence of the uniformly rotating axisymmetric Maclaurin spheroids, as parameterised by the ratio of the rotational kinetic energy to the gravitational energy $T/|W|$, exhibits a bifurcation point (at $T/|W|=0.13572$), where a new sequence of triaxial solutions branches off (Jacobi ellipsoids). Maclaurin spheroids with values of $T/|W|$ which are greater than such a threshold are {\em secularly} unstable with respect to global bar modes. Further along the Maclaurin sequence (when $T/|W|> 0.2738$), a {\em dynamical} instability with respect to a bar mode sets in. 

In the theory of rotating stars (e.g., see \citealt{1978trs..book.....T} for a comprehensive introduction), the investigation of the properties of self-gravitating rotating fluid bodies has been generalized to the case of configurations with nonuniform density, especially polytropic fluids with solid-body (e.g., see \citealp{Jam64,Sto65,Lai93}) and with differential rotation (e.g., see \citealp{OstMar68,Toh85,Hac86,Pic96,New00}). 
    
On the other hand, in the study of rotating stellar dynamical systems, stability analyses have been limited mainly to the context of disks.  \cite{Hoh71} and \cite{Kal72} provided the numerical and analytical evidence that uniformly rotating disks can be strongly unstable with respect to global bar modes, if the ordered kinetic energy dominates the energy budget of the system. \citet{1973ApJ...186..467O} extended this numerical investigation to the case of differentially rotating disks and conjectured the condition $T/|W| < 0.14 \pm 0.02$ as an {\em empirical} necessary (but not sufficient) criterion for the  dynamical stability of  any rotating stellar system with respect to global bar-like modes. 
 
Over the years, a number of studies, mostly based on N-body simulations, have been carried out to investigate the stability of stellar disks (e.g., see \citealp{Hoh76, Sel81}), so that the nature of the rotational instability with respect to bar-like modes was indeed confirmed to be dynamical. 
Few attempts at providing a physical interpretation of the \citet{1973ApJ...186..467O} criterion have been made (\citealp{Van82}); several counterexamples or alternative stability criteria have also been suggested (e.g., see \citealp{ZanHoh78,BerMar79,Aoki79,Efs82,Evans98,Ath08}).

Yet, the stability properties of differentially rotating {\em spheroidal} stellar systems have been rarely explored (see \citealp{Pal90,KuiDub94, SelVal97}) and the connection with the corresponding fluid systems is only partially understood (see \citealp{Chr95}). Some exceptions are represented by the collisionless counterparts of the uniformly rotating polytropes \citep{Van80}, Riemann ellipsoids \citep{VanWel82}, and Maclaurin spheroids \citep{Van91}, for which analytical studies have demonstrated that the dynamical instability with respect to bar mode sets in at lower values of rotation with respect to the Maclaurin fluid sequence (e.g., at $T/|W|=0.17114$, see \citealp{Chr95}). 

More recently, there has been a revival of interest in the study of the stability of differentially rotating fluids, kindled by the surprising discovery by \citet{Cen01} of an unstable $m=1$ mode in polytropes with strong differential rotation, polytropic index $n=3.33$, and $T/|W| \approx 0.14$. Since then, numerical studies have confirmed that $m=1,2$ modes can become unstable in a variety of differentially rotating fluid models, having values of $T/|W|$ as low as $0.01$ (\citealp{Shi02,Sai03,KarEri03}). The study of the stability of differentially rotating spherical shells (\citealp{Wat03, Wat04}) suggests that the unstable modes within a low-$T/|W|$ differentially rotating configuration are characterised by corotation within the system. In particular, \citet{Wat05} argued that differential rotation must be sufficiently strong in order for instability to occur. Numerical investigations of the stability of differentially rotating polytropes by \citet{Sai06} and \citet{OuToh06} strengthened this picture, which has now been formalised also in an analytical setting \citep{Sai16,Yoshida17}. 

In a broader stellar dynamical context, it has been known for over a century that corotation plays a central role in the stability of shear flows (e.g., see \citealp{Orr07,Lin55}); in the case of self-gravitating stellar disks, this intuition has offered a key guiding principle since the beginning of the development of the density wave theory of spiral structure (e.g., see \citealt{2008gady.book.....B}, \citealp{BerLin96} and references therein), especially in the context of various possible mechanisms underpinning the formation of a central bar in disk galaxies (e.g., see pioneering studies by \citealp{1972MNRAS.157....1L, Mark77, Toomre81}). 

Leaving aside for a moment the role of the angular momentum, it is worth remembering here that a number of instructive investigations have been focussed on the behaviour of \textit{spherical}, non-rotating stellar systems in a purely collisionless setting. Also in this case, the analogy with simple fluid systems has offered a fruitful pathway: a numerical stability analysis of the collisionless counterparts of classic polytropes has been conducted by \citet{Henon73}, while an analytic criterion for the stability of the case of polytropic index $n=3/2$ has been proposed by \citet{Antonov87}. Such studies are also representative examples of the two main approaches broadly adopted in this context: direct numerical integration of the relevant equations of motion or of the corresponding collisionless Boltzmann equation (e.g., see \citealp{Barnes86, Fuj83}, respectively) and variational principles in energy space (e.g., see \citealp{Lyn69,LynSan69,Doremus71,Gillon80}). More recently, the stability analysis of isotropic spheres has also been extended beyond the search for bar modes, and the possibility of the existence of lopsided, weakly decaying modes has been revealed (see \citealp{Wein94,Heg20} and references therein). 

Subsequent investigations of the behaviour of collisionless spherical equilibria have progressively ventured into the anisotropic regime, mostly by means of techniques based on the analysis of the relevant response matrix or a variety of N-body approaches. Much attention has been devoted to the case of radially biased systems (e.g., see \citealt{DejMer88,MerAgu85,PalPap87,Saha91,Ber94,Poly2007,Poly15} among many others), possibly under the stimulus offered by the identification of a global criterion for the so-called ``radial orbit instability'', first proposed by \citet{Pol81}. \alv{ On the other hand, tangentially anisotropic systems have enjoyed a lower degree of attention, with only a few studies addressing the non-trivial investigation of their stability (see especially 
{\citealt{Mikhail71,Pal74,Barnes86,Pol87};} and \citealt{Palmer+1989}). Selected studies} have also considered the case of spherically symmetric, rotating, collisionless systems, primarily with \sr{approaches based on the linear response theory (e.g., see the early works on homogeneous systems with rotation by  \citealt{Syn1971,Morozov1974})} or the \sr{direct} numerical integration of the equations of motion (e.g., see \citealt{Perez96a,Perez96b,Ali99,Mez02}). A comprehensive description of several results achieved in this area of research is offered by \sr{\citet{Pol81}} and \citet{Palmer1994}. For the purpose of the analysis conducted here, we wish to highlight in particular two mechanisms that can give rise to linear instability in, respectively, tangentially anisotropic and rotating spherical stellar systems: the ``circular orbit instability'' \citep{Palmer+1989} and the ``tumbling instability'' \citep{Allen+1992}.  

The present study offers a further step along this classic line of theoretical development, by conducting a stability analysis of several classes of equilibrium models characterised by spherical symmetry, anisotropy in the velocity space and differential rotation, or shear. The properties of the equilibria under consideration certainly can not capture, in full generality, the problem of investigating the stability of spheroidal, anisotropic, rotating stellar systems, nonetheless they offer a useful setting to bridge our fundamental understanding of axisymmetric systems with different shear prescriptions and spherical systems with various degrees and flavours of anisotropy. Particular attention is paid to the case of tangentially anisotropic systems, which, compared to the radially anisotropic ones, have been explored to a much lesser extent. The analysis of the tangentially anisotropic regime has certainly much to reveal yet, as evidenced by a new sequence of globally unstable, growing modes which has been recently identified by \citet{Rozier+}. 

This study is also part of a series of articles devoted to the study of ``kinematic richness'' (i.e., any deviations from the simplifying assumptions of isotropy in the velocity space and absence of internal rotation), which is empirically emerging in a variety of stellar systems, including star clusters. Indeed, the second Data Release of the European Space Agency mission Gaia has opened the era of ``precision astrometry'' for the study of Galactic astronomy \citep{Gaia18}. This new generation of astrometric data, complemented by proper motion measurements from decades-long campaigns with the Hubble Space Telescope \citep[e.g., see][]{AndKin03,Bel17} and state-of-the-art spectroscopic surveys \citep[e.g., see][]{Fer18,Kam18}, enable us to unlock, for the first time, the full ``phase space'' of Galactic globular star clusters in the six-dimensional position and velocity space. Such recent kinematic studies have also further confirmed the growing evidence that the presence of internal rotation in globular clusters is much more common than previously assumed \citep[e.g., see][]{Bia18,Sol19} and have enabled direct measurements of the degree of anisotropy of their velocity space \citep{Mil18,Jin19}. This rapidly evolving observational landscape, therefore, offers the motivation to address a number of old and new questions concerning the stability properties of this class of stellar systems, with a focus on the role of internal rotation and velocity anisotropy. More specifically, this contribution generalises the spherical equilibria adopted by \citet{Breen+2017} for the study of the role of primordial anisotropy in the dynamical evolution of collisional stellar systems (Part I) and prepares the ground for a further evolutionary analysis which will enrich our current fundamental understanding of the role of angular momentum in collisional stellar dynamics (Part III, Breen et al., in preparation).

The present article is organised as follows. Section 2 describes the classes of equilibria adopted as initial conditions of our N-body simulations and outlines the methodology employed for the \dch{study and} characterisation of the relevant unstable modes. The results of our stability analysis are presented in Section 3, together with a study of the spatial-temporal structure of the instability. Next, in Section 4, we investigate the role of resonances, supplementing the analysis of our N-body models with a comparative study conducted with the aid of the matrix method (as customised to rotating spheres by \citealp{Rozier+}). Finally, in Section 5 we discuss our results within existing theoretical mechanisms for bar formation and the broader context of global stability criteria for rotating, anisotropic stellar systems.

\section{{Method}}

\subsection{Matrix and $N$-body methods}\label{sec:methods}

For the study of collisionless stability in stellar systems, the two most important methods are $N$-body techniques and matrix methods.  The latter deal only with linear stability, and tend to be less flexible, but provide a direct insight into the dynamics, being naturally expressible in terms of action-angle variables (or other invariants).  We, therefore, adopt this approach in Sec.\ref{sec:matrix}, where we aim to study the role of various resonances in the observed instabilities.
In order to explore a wide range of initial conditions (especially in the rotation profile), however, the results we present first (\dch{Sections \ref{sec:prograde-results}--\ref{sec:gamma34}}) adopt particle methods. \dch{The following paragraphs consider these two kinds of techniques in turn.}

Generally, we avoid direct-summation $N$-body methods, because the need to complete many simulations in a reasonable time tends to restrict systems to modest $N$, when stability may be modified by collisional processes. It has even  been said that they  ``should be regarded as methods of last resort'' \citep{1997ASPC..123..215S}.  Therefore we mainly adopt an efficient (hierarchical) method which makes it feasible to carry out simulations with larger particle numbers, and employs softening to further reduce collisionality; for these reasons our method of first resort in this work is \texttt{gyrfalcON} \citep{2002JCoPh.179...27D}.  This is used to generate the main bulk of the numerical results, which are in the earlier parts of Sec.\ref{sec:results}.  Elsewhere, however, in Sections \ref{sec:spatiotemporalstructure} and \ref{sec:nbody-resonances}, we do resort to use of the collisional  code \texttt{NBODY6} \citep{2012MNRAS.424..545N}.

\dch{The matrix method dates back to the work of \citealp{Kal77} (though his emphasis was on disks) and to the study of non-rotating spherical systems by \citet{Pol81}.  For such systems the formulation we employ \citep[][see, for example, \alv{E}q.(\ref{eq:RespMat}) in the present paper]{Rozier+} would be identical to that in the Appendix of \citet{Pol81}, except for notation, which in our case is derived from  \citet{Ham18}.  
Our choice of basis functions (of biorthogonal density-potential pairs) can be found in \citet{HernOst92} and relies on ultra\-spherical polynomials.
In the case of rotating systems our formulation of the matrix equation may also differ a little from those found in the literature \citep[e.g.][\alv{e}q.(8.18)]{Palmer1994}, not only because of notation, but also through the choice of distribution function, though such differences are minor.  
 }

Throughout the paper (even in sections not devoted to $N$-body simulations), units are such that $G$, the total mass and virial radius are unity (i.e. H\'enon Units).

\subsection{Initial conditions}\label{sec:ics}

The initial conditions we adopt all have the same density profile, that of a Plummer model, and the velocity distribution is constructed in a series of steps:
\begin{enumerate}
    \item We begin with a model from the series constructed by \citet{Dejonghe1987}.  These are parametrised by a single dimensionless parameter, $q$, in the range $(-\infty<q\le2)$.  Increasing positive values correspond to models with increasing radial anisotropy, while increasingly negative values correspond to those with increasingly tangential anisotropy.  The Einstein-Plummer model, with all stars on circular orbits, corresponds to the limit $q\to-\infty$.
    \item The velocities are adjusted so that a fraction $\alpha$ of the particles with negative azimuthal velocity $\vphi$ (in spherical coordinates $r,\theta,\phi$) are changed from negative to positive in a certain region of phase space, as described below.  Thus $\alpha = 0$ corresponds to Dejonghe's model, and when $\alpha=1$ all particles have positive azimuthal velocity.  We refer to this change in the velocities as the action of ``Lynden-Bell's demon" \citep{1962MNRAS.123..447L}, and we use the terms ``prograde" and ``retrograde" to mean that $\vphi>0$ or $<0$, respectively.  (The demon can be applied in reverse, as in (b) below.)
    \begin{enumerate}
        \item {\sl Basic models:} In the simplest series of models, the demon is allowed to work only on a certain fraction of the mass ($0\le\gamma_1\le1$) when the particles are ordered by energy. {Thus $\gamma_1 = 0.2$ means that the Lynden-Bell demon works on the 20\% most bound particles; $\gamma_1$ = 0 gives a Plummer-Dejonghe model, while when $\gamma_1=1$ and $\alpha=1$, all particles are given prograde rotation.}  These models are intended to give insight into the range of radii which are active in the  instabilities observed in them.
        \item {\sl High-shear models:} In the next series of models, the above procedure is applied to all particles (or rather to a fraction $\alpha$ of them) in the most bound fraction $\gamma_2$, but in the rest, i.e. the least bound fraction $1-\gamma_2$ of them, the Lynden-Bell demon acts in reverse, reversing  $\vphi$ for all prograde stars (or a fraction $\alpha$ of them).  These models are intended to maximise the shear in the rotation profile  (Sec.\ref{sec:analysis}), as this may be a factor in the existence of instabilities.
        \item {\sl Low-inclination models:} Third, we consider a series of models in which the demon works only on stars of low inclination $i<\gamma_3\pi/2$, where $i$ is defined in the sense of the inclination of the orbital plane to the equatorial plane $z=0$, so that $0\le i\le\pi/2$.
        A motivation here is to investigate to what extent an instability can be thought of as being driven by motions in a thick conical disk.
        \item {\sl High-inclination models:} Lastly \dch{we have} a set of models in which the Lynden-Bell demon works only on stars of high inclination $i > \gamma_4\pi/2$.   Such models may give additional indications of the geometrical and kinematic characteristics of orbits which contribute strongly to instability.
    \end{enumerate}
\end{enumerate}

These are the only initial models considered in this paper, but many variants are possible.
For example, in the high-shear models, the axis of rotation of the least bound stars may be chosen differently from that of the most bound stars.  Many such models can be explored by setting up initial conditions with the software \texttt{PlummerPlus}\footnote{https://github.com/pgbreen/PlummerPlus}.

\subsection{Characterisation of rotation}\label{sec:analysis}

There are several ways in which the rotation of these models can be measured.
The first is a rotation profile, which can be defined in different ways, e.g. in spherical or cylindrical geometry, of which we concentrate on the latter.  It is easily calculated from any  model by first calculating the average rotation speed $\vphibar(R) = \langle\vphi\rangle$ as the average azimuthal velocity in a cylindrical shell of cylindrical radius $R$, and a thickness chosen as a compromise between sufficient spatial resolution and sufficient statistical reliability.  From an $N$-body simulation the corresponding quantity is easy to compute on a grid of $R$.  Several examples are given below (Sec.\ref{sec:profiles}).

For a global measure of rotation we compute the rotational kinetic energy, but for this purpose define the average rotation speed on a circular ring of given $R$ and $z$ and, again, suitably chosen bin size.  Then the rotational kinetic energy is defined to be 
\begin{equation}
  T = \frac{1}{2}\int \rho(R,z) \vphibar^2(R,z) 2\pi R dR dz,
\end{equation}
the integral being evaluated in practice by summing over the chosen grid.  For our purpose a fixed grid in $R$ with 100 bins, and a grid in $z$ adapted to contain approximately equal numbers of particles, was adopted.   Of particular importance in this field, because of the pioneering work of \citet{1973ApJ...186..467O}\alv{,} is the dimensionless parameter $T/W$, where $W$ is the binding energy of the system. 

\subsection{Rotation profiles}\label{sec:profiles}

\begin{figure}
    \centering
        \includegraphics[width=0.5\textwidth]{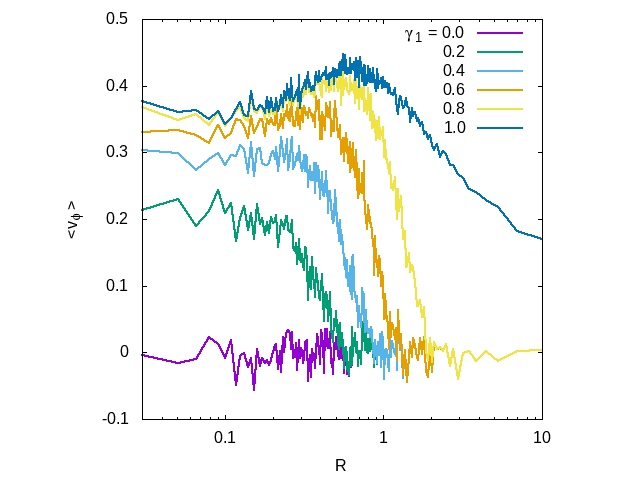}
    \caption{Rotation curves of six ``basic" models with $q = -6, \alpha=1$ and values of $\gamma_1 = 0(0.2)1$.  Here $\gamma_1$ is the fraction of stars, ordered by energy, below which the Lynden-Bell demon is applied.  The ordinate is the mean azimuthal velocity in a cylindrical shell of radius $R$ (the abscissa).  Each of the 200 shells contains 500 particles. The six plots correspond to application of the Lynden-Bell demon (with various values of $\gamma_1$) to the same realisation of a Plummer-Dejonghe model; therefore the fluctuations in successive models are correlated.}
    \label{fig:rot-curves}
\end{figure}

Fig.\ref{fig:rot-curves} illustrates various rotation curves 
for the set of ``basic" models (Sec.\ref{sec:ics}).
In this figure the uppermost (blue) curve shows the rotation curve of a model where the Lynden-Bell demon is applied to all stars, whereas the lowest shows the rotation curve of the original Plummer model (anisotropy in all the models being modified by setting $q = -6$), and this is non-zero only because the profile has been determined from an $N$-body model.  The green curve, corresponding to $\gamma = 0.2$, shows the result when the demon is applied only to the 20\% most bound particles. 
Other choices of $\alpha$ simply alter the ordinate in proportion to $\alpha$.  

\begin{figure}
    \centering
        \includegraphics[width=0.5\textwidth]{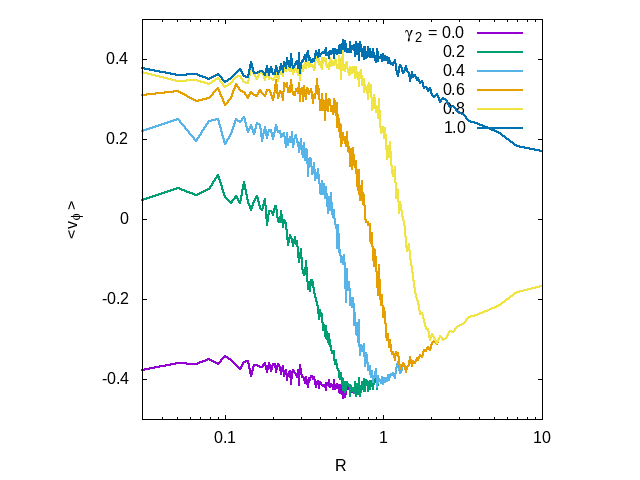}
    \caption{Rotation curves of six ``high-shear" models with $q = -6, \alpha=1$ and values of $\gamma_2 = 0(0.2)1$.  As Fig.\ref{fig:rot-curves}, except that here $\gamma_2$ is the fraction of stars, ordered by energy, below which the Lynden-Bell demon is applied in the sense of prograde rotation, while above that energy it is applied in such a way as to give retrograde rotation.  }
    \label{fig:rot-curves-g1}
\end{figure}

Fig.\ref{fig:rot-curves-g1} gives similar plots for the ``high-shear" (counter-rotating) models, in which $\gamma_2$ is the fraction of stars, ordered by energy, below which the Lynden-Bell demon is applied to give prograde rotation, while it is applied to the less bound stars to give counter-rotation.  Qualitatively the rotation curves look very similar, but the differences between the maximum and minimum (largest retrograde) velocities are about twice as large.

\begin{figure}
    \centering
            \includegraphics[width=0.5\textwidth]{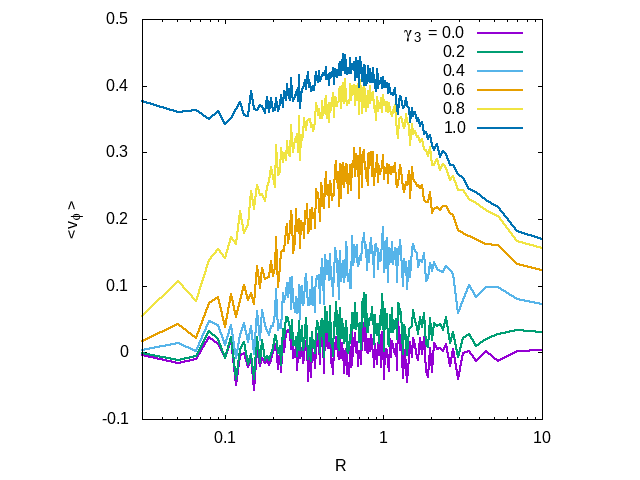}
    \caption{Rotation curves of six ``low-inclination" models with $q = -6, \alpha=1$ and values of $\gamma_3 = 0(0.2)1$.  As described in the text, the Lynden-Bell demon is applied only to low-inclination orbits, with $i < \gamma_3\pi/2$.  The ordinate is the mean azimuthal velocity in a cylindrical shell of radius $R$ (the abscissa).  Each of the 200 shells contains 500 particles.}
    \label{fig:rot-curves-g3}
\end{figure}

\begin{figure}
    \centering
    \includegraphics[width=0.5\textwidth]{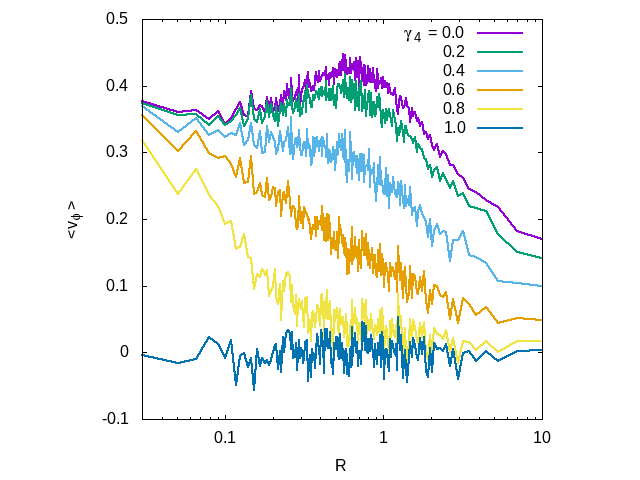}
    \caption{Rotation curves of six ``high-inclination" models with $q = -6, \alpha=1$ and values of $\gamma_4 = 0(0.2)1$.  The Lynden-Bell demon is applied only to high-inclination orbits, with $i > \gamma_4\pi/2$.  The ordinate is the mean azimuthal velocity in a cylindrical shell of radius $R$ (the abscissa).  Each shell contains 500 particles.}
    \label{fig:rot-curves-g4}
\end{figure}

Next, we can consider the effect of the inclination of the rotating orbits, by applying the Lynden-Bell demon  only to orbits of low inclination; specifically, to stars  such that $i < \gamma_3\pi/2$.   The resulting rotation curves are illustrated in Fig.\ref{fig:rot-curves-g3}.  At small cylindrical radius this procedure greatly reduces the amount of rotation (unless $\gamma_3$ is very large), since most stars at small cylindrical radius must lie on high-inclination orbits.  The effect at large radius is much smaller, unless the critical value $\gamma_3$ is very small.  In a complementary fashion we can also choose to apply the demon only to high-inclination orbits, such that $i > \gamma_4\pi/2$, which leads to the rotation curves in Fig.\ref{fig:rot-curves-g4}.

\subsection{Analysis of results}

Our method of searching for a bar, in a spherical model which is rotating about the $z-$axis, is described in \citet{Rozier+}, along with studies of its accuracy\footnote{See also the caption to Fig.\ref{fig:m2d-v-op}.}.  Briefly, a bar mode in the $x,y$ plane is measured by computing the following integral, which uses cylindrical polar coordinates:
\begin{equation}
C_{m} (t) = \!\! \int_{0}^{\infty} \!\!\!\!\!\! R\rd R \!\! \int_{- \infty}^{\infty} \!\!\!\!\!\! \rd z \, \!\! \int_{0}^{2 \pi} \!\! \frac{\rd \phi}{2 \pi} \, \rho (R , z , \phi , t) \, \re^{- \ri m \phi} ,
\label{def_Cm}
\end{equation}
where $m=2$ for a bar, and the integral is replaced by a sum of delta functions, one for each particle.
Then the growth rate is computed by fitting an exponential\footnote{A logistic fit is used in Appendix \ref{sec:method-of-analysis}.} to $\vert C_m(t)\vert$ over a suitable time interval.  Actually this is a relatively insensitive measure for non-rotating systems ($\alpha = 0$ in this work), as the bar (if there is one) need not form in the $x,y$ plane.  Therefore non-rotating systems ($\alpha = 0$) can be measured by computing the corresponding integrals involving spherical harmonics of order $l=2$, but 
\sr{this is not what is done in the present study.}

\section{Results}\label{sec:results}

\subsection{Basic models (prograde rotation)}\label{sec:prograde-results}

\begin{figure}
    \centering
        \includegraphics[width=0.5\textwidth]{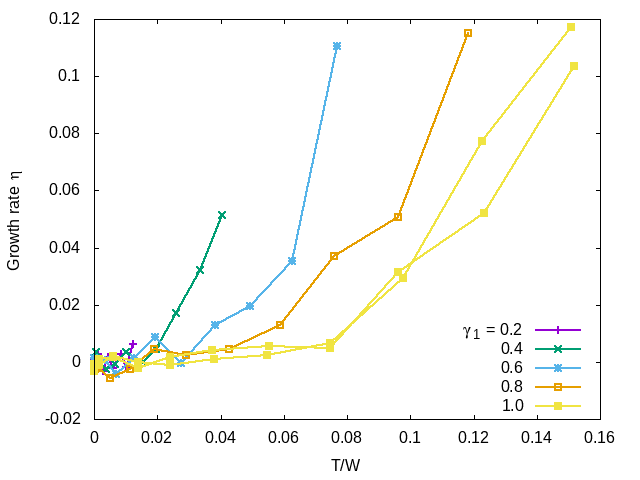}
    \caption{Growth rate versus Ostriker-Peebles parameter for six sets of models with $q = -6, 0\le\alpha\le 1$ and various $\gamma_1$.   Note that two realisations have been computed for the case $\gamma_1=1$, which gives an impression of the accuracy with which the growth rate has been estimated.}
    \label{fig:m2d-v-op}
\end{figure}

We begin (Sec.\ref{sec:ics}) with the models in which prograde rotation is added at energies less than (more bound than) a certain fraction $\gamma_1$ of mass.  Let us consider first the fiducial case $q = -6$. Fig.\ref{fig:m2d-v-op} shows the growth rate $\eta$ as a function of the Ostriker-Peebles parameter $T/W$, where $T$ is the rotational (mean flow) part of the kinetic energy. Note that results for $\gamma_1 = 0$ are omitted, as these models are stable \dch{(in the sense that $\eta$ is not significantly non-zero)}, according to  \citet{Rozier+}.  Results for $\gamma_1 = 1$ are duplicated, i.e. two independent realisations are shown for each value of $\alpha$, to give an indication of the numerical uncertainty in the measured growth rate.  Each sequence of symbols corresponds to a given value of $\gamma_1$, and one moves to the right within a given sequence as $\alpha$ increases.  This increase in $\eta$ is satisfactorily fitted by a \dch{cubic power law in $T/W$, or equivalently}
\begin{equation}
    \eta(q,\gamma_1,\alpha) = \mumax(q,\gamma_1)\alpha^6,
\end{equation}
a remarkably steep dependence. 

\begin{figure}
    \centering
        \includegraphics[width=0.5\textwidth]{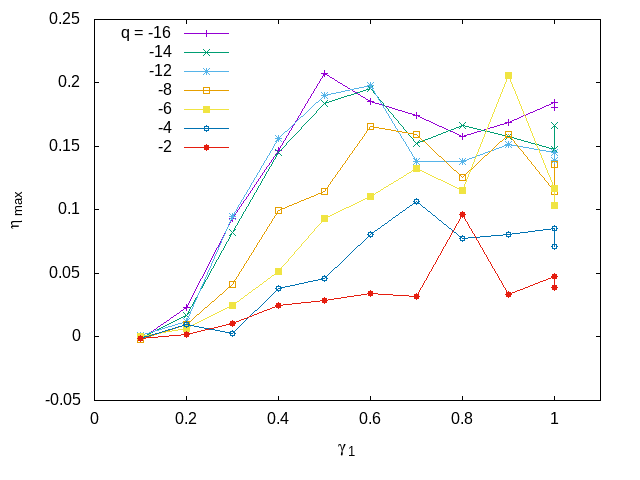}
    \caption{Growth rate for the maximally rotating model at each $\gamma_1$, for several values of $q$.  We have avoided the isotropic case $q = 0$, as it is known \citep{Rozier+} that \sr
    {any} instabilities here 
    \sr
    {could} belong to a different family from those at smaller (more negative) values of $q$.  The result of two independent simulations is given for $\gamma_1 = 1$.} 
    \label{fig:mumax}
\end{figure}

Because of the scaling with $T/W$, we shall now only consider the function $\mumax(q,\gamma_1)$, results for which are shown in Fig.\ref{fig:mumax}.  Results of two models are given at $\gamma_1=1$, again to give an impression of the reproducibility of the results for independent initial conditions.  Even the large fluctuations (such as the maximum at $\gamma_1=0.8$ in the lowest curve) may be explicable in terms of the choice of initial conditions, but trends in the data cannot be so discounted.  The qualitative point that one may take from this plot is the fact that the growth rate does not keep increasing as the fraction $\gamma_1$ increases.  Rather, the growth rate saturates when roughly the inner half of the system (ordered by energy) has been set to rotate in one sense\footnote{Actually there is some evidence from this figure that the saturation level is a little lower than the peak level, as one notices particularly for the most negative values of $q$.\label{footnote}}.  Turning back to Fig.\ref{fig:rot-curves}, in particular the rotation curves for $\gamma = 0.6, 0.8$ and $1$, the large differences in these curves, which are particularly evident outside $r \sim 0.5$, make little difference to the growth rate. By contrast the rotation curves for $\gamma_1 = 0.2$ and $0.4$ change dramatically inside $r\sim 0.5$, as do the growth rates (Fig.\ref{fig:mumax}).  This \sr
{region ($r \ltorder 0.5$)}, then, appears to be the region within which one must seek the mechanism for the instability, and any quantitative, phenomenological model for it.

\begin{figure}
    \centering
        \includegraphics[width=0.5\textwidth]{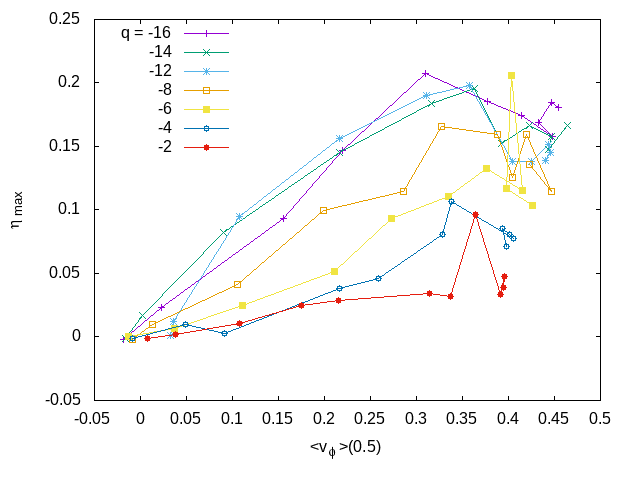}
    \caption{Growth rate for the maximally rotating model at each $\gamma_1$, for several values of $q$, plotted against the value of the rotation speed at radius $R = 0.5$.} 
    \label{fig:maxmuv5}
\end{figure}

To make this discussion more quantitative, in Fig.\ref{fig:maxmuv5} we show how the maximum growth rate (as plotted in Fig.\ref{fig:mumax}) varies with the value $\vphibar(0.5)$.  This plot shows that  this value of $\vphibar$ is a roughly linear predictor of the growth rate, though for very negative values of $q$ it does not predict well the saturation of the growth rate at the largest values of $\vphibar(0.5)$.  Incidentally, this value was conveniently available, but values at other neighbouring radii might well be equally useful, as illustrated just below.

\begin{figure}
    \centering
        \includegraphics[width=0.5\textwidth]{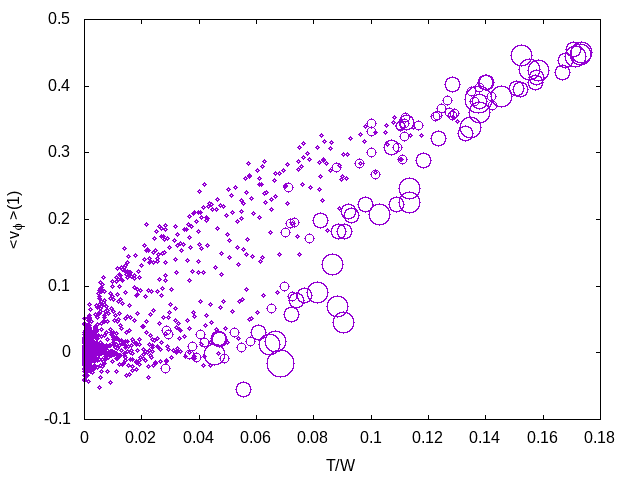}
    \caption{Scatterplot of all results \dch{on models with prograde rotation}, in the plane of $T/W, \vphibar(1)$, i.e. the Ostriker-Peebles parameter and the azimuthal speed at the virial radius.  The growth rate is coded in the size of the plotted points; the smallest symbols indicate growth rates less than 0.05.} 
    \label{fig:opv1}
\end{figure}

The maximum growth rate occurs when $\alpha=1$, i.e. the Lynden-Bell demon acts at its maximum level.  But now the entire dataset (all values of $q\le0, \alpha$ and $\gamma_1$; see Fig.\ref{fig:opv1}) can be grasped in outline by means of the foregoing analysis.  The axes correspond to the Ostriker-Peebles stability parameter and the azimuthal streaming speed at the virial radius, and the size of the symbol encodes the growth rate of the bar.  The region of the diagram containing the fastest growth rate is particularly interesting:  although the classical Ostriker-Peebles criterion (for {\sl flattened} systems) $T/W \gtorder 0.14$ corresponds rather well with the boundary of this region when $\vphibar(1)$ is large, we now find strongly unstable models when $T/W$ is much smaller, provided that $\vphibar(1)$ is small.  These are models in which $\gamma_1$ is sufficiently small that the azimuthal speed has dropped almost to zero by the virial radius (cf. Fig.\ref{fig:rot-curves}).  In fact the evident, roughly horizontal groupings in this diagram correspond to the two or three highest values of $\gamma_1$ included in our numerical survey.

\subsection{High-shear models (mixed pro- and retrograde rotation)}\label{sec:high-shear-results}

\begin{figure}
    \centering
        \includegraphics[width=0.5\textwidth]{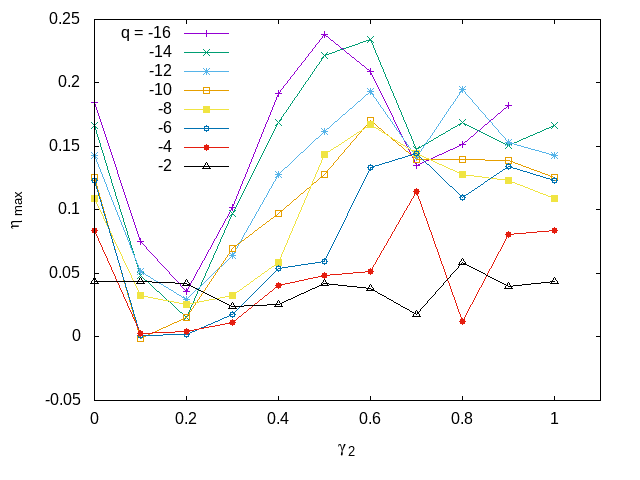}
    \caption{Growth rate for the maximally rotating (and counter-rotating) model at each $\gamma_2$, for several values of $q$.  For the reason stated in the text, the results for $\gamma_2 = 1$ have been duplicated for $\gamma_2 = 0$ (though new realisations would have led to slightly different results).} 
    \label{fig:mumax1}
\end{figure}

For this sequence of models, we assume 
that the scaling with $T/W$ is similar to that in the purely prograde case, and proceed directly to the results for $\alpha = 1$.  The growth rate is shown in Fig.\ref{fig:mumax1}.  Qualitatively the results behave quite similarly to those for the purely prograde models, and even the maximum growth rates (which occur for models with smaller, i.e. more negative, values of $q$, and then at intermediate values of $\gamma_2\simeq0.5$) are not as dissimilar as the more extreme gradient in the velocity curve (compare Figs.\ref{fig:rot-curves}, \ref{fig:rot-curves-g1}) might lead one to expect. Nevertheless one notices even more strongly that the maximum growth rate occurs at intermediate values of $\gamma_2$ (cf. footnote \ref{footnote}); furthermore the growth rate there is even larger than in Fig.\ref{fig:mumax}.  These remarks suggest that the maximum growth rate depends on the rotation of the outer (least bound) parts of the system:  as $\gamma_1$ increases beyond 0.5, these parts start to rotate, and the growth rate is depressed slightly, while as $\gamma_2$ increases beyond 0.5 (Fig.\ref{fig:mumax1}) the outer regions switch from retro- to prograde rotation, and the drop in growth rate is more pronounced and starts from a higher maximum.

In Fig.\ref{fig:mumax1} there are significant changes in the rate of growth at the left, around $\gamma_2 = 0.2$, where the values of $\mumax$ have a very pronounced minimum,  increasing again as $\gamma_2$ decreases further.  In fact for $\gamma_2=0$ the growth rates must be the same (within the error of the estimates) as for $\gamma_2 = 1$, as these two cases have the same rotation curve except for the sense of rotation.

\begin{figure}
    \centering
        \includegraphics[width=0.5\textwidth]{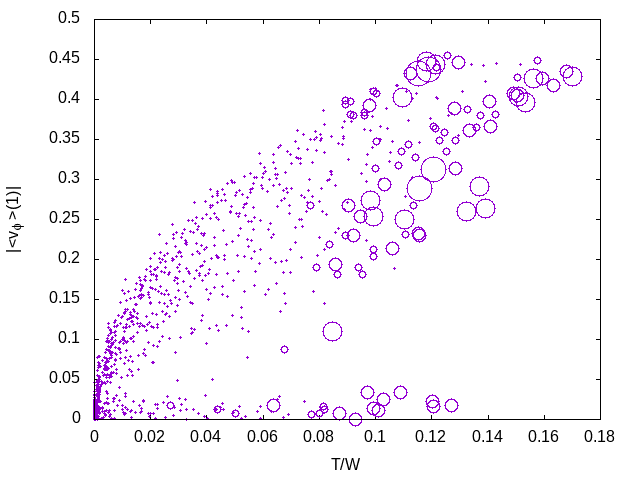}
    \caption{Scatterplot of all results on models with pro- and retrograde rotation, in the plane of $T/W, \vert\vphibar(1)\vert$, i.e. the Ostriker-Peebles parameter and the {\sl absolute value} of azimuthal speed at the virial radius.  The growth rate is coded in the size of the plotted points; the smallest symbols indicate growth rates less than 0.05.} 
    \label{fig:opv1-gamma1}
\end{figure}

If we contrast the great difference in $\eta_{max}$ between $\gamma_2 = 0$ and $\gamma_2=0.2$ with the small change in the rotation velocity at $R = 1$, it is clear that
 $\vphibar(1)$ 
 \sr
 {(coupled with $T/W$)} cannot be as simple a predictor of the growth rate as in the case of the prograde models (Fig.\ref{fig:opv1}).  In Fig.\ref{fig:opv1-gamma1} we show all results for the counter-rotating models, except that now the ordinate is the absolute value of $\vphibar(1)$.  The result is qualitatively similar to that for the prograde models, but more noticeable here is the fact that there are some very rapidly rotating models (high $T/W$ and $\vert \vphibar(1)\vert$) with relatively low growth rates.  We now investigate the rotation curves of such models to try to identify distinctive features of their rotation curves. 

\begin{figure}
    \centering
        \includegraphics[width=0.5\textwidth]{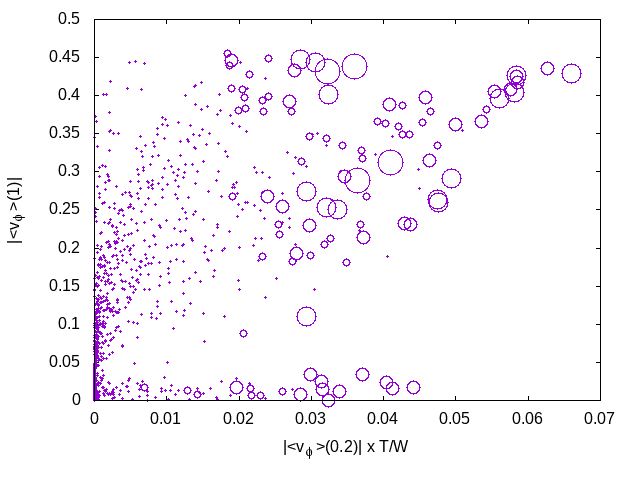}
    \caption{As Fig.\ref{fig:opv1-gamma1}, except that the abscissa is multiplied by the mean azimuthal speed at radius $R = 0.2$, to separate models which are rotating relatively slowly at small radii.  The growth rate is coded in the size of the plotted points; the smallest symbols indicate growth rates less than 0.05.} 
    \label{fig:opv.2v1-gamma1}
\end{figure}

In fact the models with low growth rate, high $T/W$ and high $\vert \vphibar(1)\vert$ tend to have small (very negative) $q$ and $\gamma_2\sim 0.2$.  Such models are also very distinctive in Fig.\ref{fig:mumax1}, as discussed above.  What distinguishes the rotation curve for these models, as can be seen in Fig.\ref{fig:rot-curves-g1}, is that the interior radii of a $\gamma_2 = 0.2$ model are hardly rotating, at least by comparison with radii above about 0.5.  We can conjecture that the growth rate is small when the rotation speed $\vphibar(0.2)$ is small.  Therefore, to separate such models better than in Fig.\ref{fig:opv1-gamma1}, we simply multiply the abscissa by $\vert\vphibar(0.2)\vert$.  The result is shown in Fig.\ref{fig:opv.2v1-gamma1}.  There is now a much better separation of slowly and rapidly growing modes, and this choice of parameters has comparable success in the basic models of Sec.\ref{sec:prograde-results}.  These results confirm our finding (in that subsection) that the kinematics of the region inside $R\simeq0.5$ is of particular importance for the presence of instability.  

While this discussion has been focused on the growth rate, it is also interesting to measure the pattern speed in relation to the coexistence of pro- and counter-rotation at different radii.  The model with $\gamma_2 = 0$ is completely counter-rotating (Fig.\ref{fig:rot-curves-g1}), and for $q=-16$ (which we focus on in this paragraph) the pattern speed is also negative: $\Omega_b\simeq-0.7$\footnote{Measurements reported in this paragraph were taken in a time interval around $7<t<21$, from the time of the clear emergence of the bar up to a time which is about half of the time when it reaches its maximum amplitude. The values are thought to be accurate to better than 10\%.}. The pattern speed reduces in magnitude to -0.5 at $\gamma_2 = 0.1$, but is about +0.7 (i.e. prograde) throughout the range $0.3\le \gamma_2\le1$.  At $\gamma_2 = 0.2$ the instability is complicated, and no pattern speed could be measured confidently: up until $t\simeq40$ a weak bar grows with a fast pattern speed around $\Omega_b = -1.7$, but this gives way by  $t\simeq60$ to a bar with faster growth and $\Omega_b\simeq-0.4$.

\subsection{Models with prograde rotation at low or high inclination.}\label{sec:gamma34}

\begin{figure}
    \centering
        \includegraphics[width=0.5\textwidth]{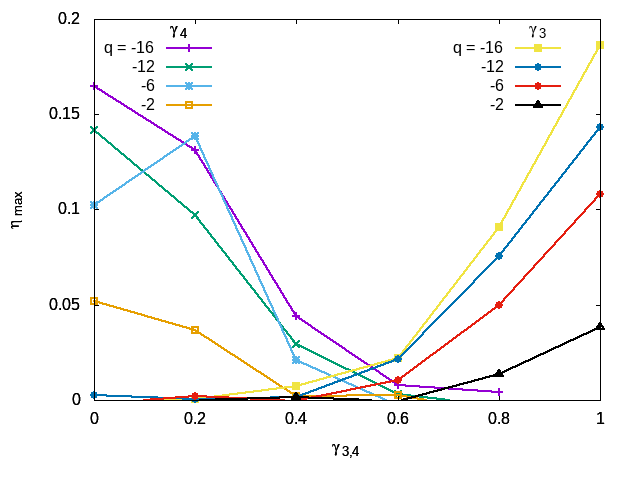}
    \caption{As Fig.\ref{fig:mumax1}, except that the Lynden-Bell demon is applied at only low or high inclinations, as determined by the parameters $\gamma_3$ and $\gamma_4$ respectively.} 
    \label{fig:mumax34}
\end{figure}

More briefly, we consider the maximal growth rate (i.e. for $\alpha=1$) for models in which rotation is given only to orbits of low inclination, i.e. those with rotation curves exemplified by Fig.\ref{fig:rot-curves-g3}, or only to orbits of high inclination (Fig.\ref{fig:rot-curves-g4}).  The results are shown in Fig.\ref{fig:mumax34}, with the low-inclination series on the right.  As $\gamma_3$ decreases from 1 to about 0.8, the growth rate drops by about a factor of 2, showing that stars with inclination above about $70^\circ$ are very important in the instability.  (This does not mean that stars of low inclination are irrelevant, but they are not sufficient of themselves for rapidly growing instability.)  A complementary conclusion is drawn from the high-inclination models, whose growth rate drops by a comparable factor if $\gamma_4$ increases from 0 to about 0.3, i.e. if one excludes orbits with inclination less than about $30^\circ$.  Thus the presence of both high- and low-inclination stars is necessary for rapid growth, but neither group by itself is sufficient. 

\subsection{Spatial-temporal structure of the instability}\label{sec:spatiotemporalstructure}

\subsubsection{Spatial structure}
\label{sec:spatialstructure}

In this subsection we focus on the basic case  $q = -16, \alpha = \gamma_1 = 1$, where the effects of rotation are strong, and apparently representative of all the very tangentially anisotropic models we have studied, including the Einstein sphere ($q = -\infty$).  Our aim is to present briefly some information on the spatial extent and shape of the bar.  For other values of $q$, representations of the density are shown in \citet[][figs.5,6 for $q = -6$, and fig.10 for other $q<0$]{Rozier+}.

\begin{figure}
    \centering
        \includegraphics[width=0.5\textwidth]{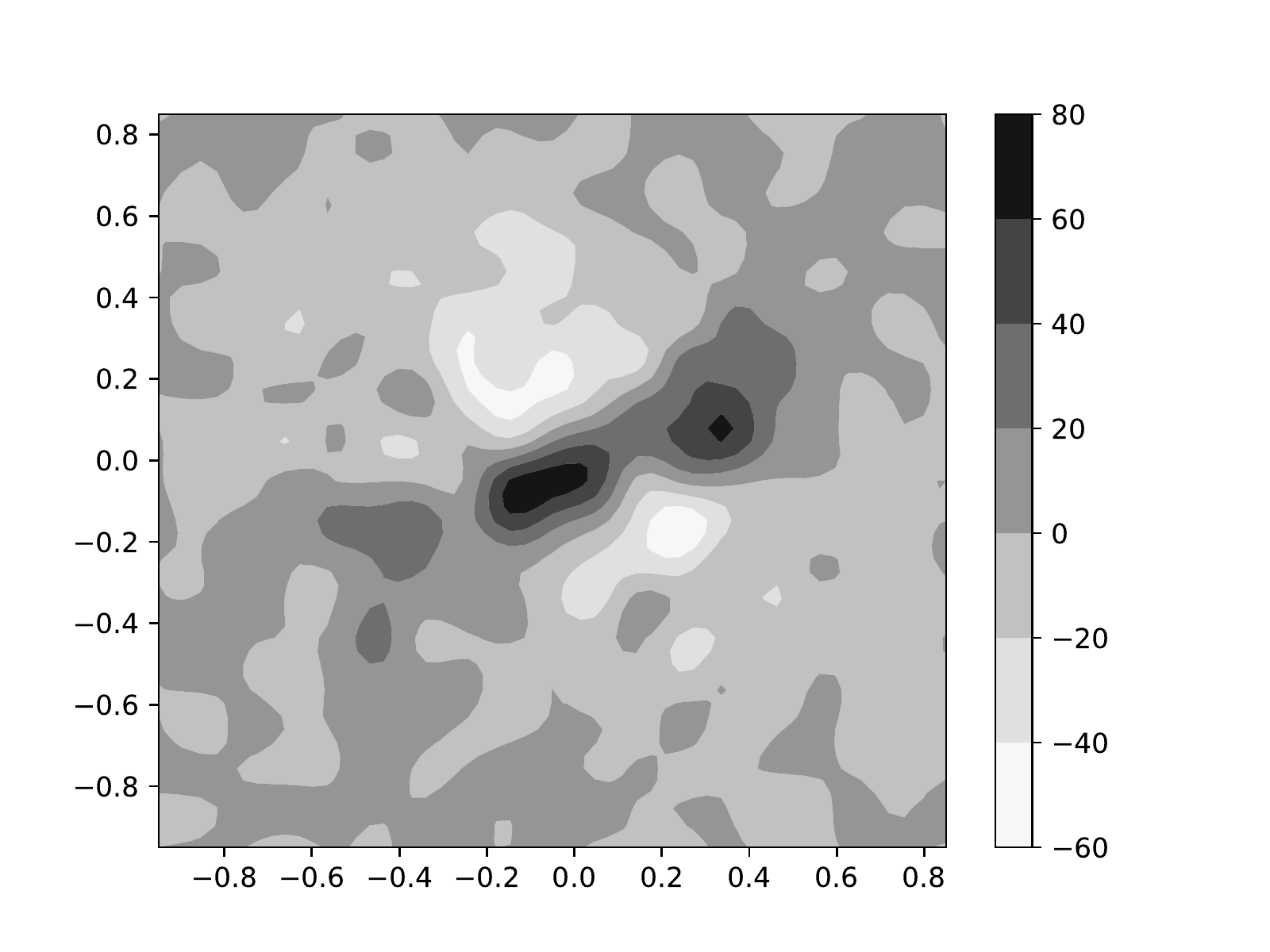}
    \caption{Projected density, in the plane $z=0$ of the bar in an $N$-body model with $N=16384$ at $t = 20$HU.  The result has been obtained by subtracting the projected density at $t=0$, and therefore shows only the perturbation.  The data was gridded with a pixel size $0.2^2$HU, and  contours were smoothed by interpolating to a grid 9 times finer.  The scale gives the number per pixel.}
    \label{fig:nbodyprojection}
\end{figure}

In $N$-body simulations with $N=16384$, which were conducted using \texttt{NBODY6}, the bar is detectable (by Fourier analysis) only inside the cylinder  $R\ltorder1.5, \abs{z}\ltorder0.9$.  Fig.\ref{fig:nbodyprojection} displays the projected density in the plane $z=0$ at time $t=20$ (in this model the bar reaches  its maximum amplitude at about $t=40$). This can be compared with the result in \citet[][the uppermost inset in their fig.10]{Rozier+}, obtained by the matrix method (Sec.\ref{sec:method}). The maximum projected radius in Fig.\ref{fig:nbodyprojection} (at the corners) is nearly the same as the maximum cylindrical radius quoted above, and faint extensions can be seen at these radii.  Also of interest here are the parts at right angles to the bar with negative projected density.  In the $N$-body model at time $t=20$, the core radius has decreased from an initial value of about 0.32HU to a value of about 0.26HU, presumably because of early core collapse.  Possibly for this reason, the projected density at the centre of the bar is above zero.

{}

\subsubsection{Growth rate and pattern speed}\label{sec:temporalstructure}\label{SEC:TEMPORALSTRUCTURE}

\begin{figure}
    \centering
    \includegraphics[width=0.5\textwidth]{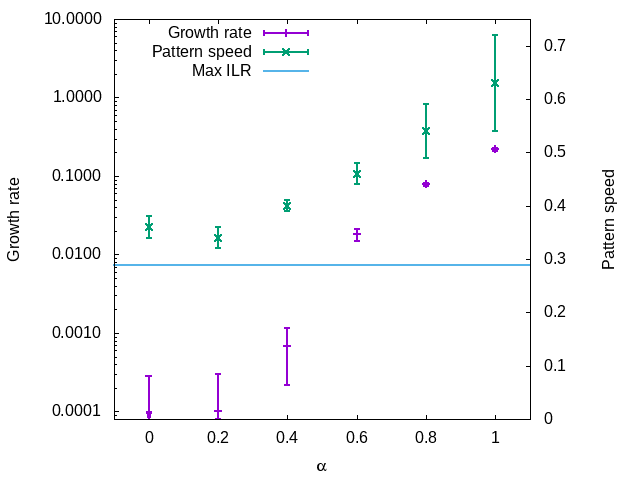}    
    \caption{Growth rate and pattern speed of direct $N$-body models with $N = 32768, q=-16,\gamma_1=1$ for various values of $\alpha$.  For $\alpha=0,0.2$ results are obtained by combining 10 models each, for $\alpha=0.4$ there were 8, and there was one model each for the three largest values of $\alpha$. For $\alpha=0$ the confidence interval on the growth rate extends to negative values, \dch{and only the upper limit is shown}.  Also shown, to illustrate the theoretical discussion of Sec.\ref{sec:mechanisms}, is the maximum of $\Omega_2 - \Omega_1/2$ for nearly circular orbits, which is related to the inner Lindblad resonance (ILR). }
    \label{fig:temporalstructure}
\end{figure}

The results that follow are based on a dedicated set of direct $N$-body models with $N=32768$ particles, with anisotropy parameter $q=-16$. They are basic models of the type discussed in Sec.\ref{sec:prograde-results} with $\gamma_1=1$, i.e. the Lynden-Bell demon is applied everywhere. Our aim is to discuss the dependence of the time scales of the bar instability on the rotation parameter $\alpha$, with a view to informing the discussion of Section \ref{sec:empirical-evidence}.  Detailed discussion of the methods of analysing the runs are given in Appendix \ref{sec:method-of-analysis}.  Results are given in Fig.\ref{fig:temporalstructure}.

{}

While details of the analysis and interpretation of this plot are given in later sections and the appendix, explanation is required here for the fact that any pattern speed at all has been measured for the non-rotating case $\alpha=0$.  \dch{We emphasise that} no growth 
could be measured with confidence; all that is shown \dch{in Fig.\ref{fig:temporalstructure}} is an upper limit.  But the analysis of the $m=2$ Fourier coefficient $C_2$ (see Eq.(\ref{def_Cm})), using the autocorrelation of $\arg(C_2)$, clearly demonstrated the presence of oscillation with the stated pattern speed.  The figure supports the view that the bar instability in models with $\alpha>0$ connects continuously with the dynamical behaviour of the non-rotating model, \dch{but does not imply that the non-rotating model is unstable}.

\section{Resonances}

\subsection{Resonances in tangentially biased spherical systems}\label{sec:resonances-in-systems}

The role of resonances in the occurrence of bar instabilities in disks is sufficiently well known to be textbook material \citep[e.g.][]{2008gady.book.....B}.  Less \dch{familiar}, however, is their role in spherical systems, despite the substantial literature referred to in Sections \ref{sec:introduction} and \ref{sec:methods}.
In this introductory subsection we focus on (very) tangentially anisotropic systems, where the radial motions are small, and we can employ the epicyclic approximation.  Then the main resonances can be identified by a quite elementary calculation (Appendix \ref{app:perturbation}); it leads to the results given in Table \ref{tab:resonances}, which we now explain.

\begin{table}
    \centering
    \caption{Resonances in the leading-order perturbations of epicyclic motions in a rotating spherical model.  Names which are not in common use are given in slanted type.  Here the orbital inclination $i$ is used in its conventional sense, with retrograde orbits having $\pi/2<i\le\pi$.}
    \begin{tabular}{c|c|c|c|}
    $\Omega_b =$    &   $i$-dependence  &   Name    & $(n_1,n_2)$\\
    \hline
    0&$\sin^2i$&{\sl tumbling}  &(0,0)\\
$\Omega$  &  $(1+\cos i)^2$   &   corotation  & $(0,2)$\\
$\Omega - \kappa/2$         & $(1+\cos i)^2$   & ILR &$(-1,2)$\\
$\Omega + \kappa/2$         & $(1+\cos i)^2$   & OLR &$(1,2)$\\
 $\pm \kappa/2$ & $\sin^2i$ &   {\sl epicyclic} &$(\pm\dch{1},0)$\\
 $-\Omega$  &  $(1-\cos i)^2$   &   {\sl reverse corotation}    &$(0,-2)$\\
$-\Omega + \kappa/2$         & $(1-\cos i)^2$   & {\sl reverse ILR} &$(1,-2)$\\
$-\Omega - \kappa/2$         & $(1-\cos i)^2$   & {\sl reverse OLR} &$(-1,-2)$
    \end{tabular}
    \label{tab:resonances}
\end{table}

We suppose there is a bar perturbation, whose pattern rotates with angular speed $\Omega_b$.  Then the familiar corotation resonance (line 2) is given by $\Omega_b = \Omega$, where $\Omega$ is the circular angular velocity.  Similarly the Inner and Outer Lindblad Resonances (ILR, line 3; OLR, line 4) are defined by $\Omega_b = \Omega \mp \kappa/2$, 
where $\kappa$ is the epicyclic frequency.  What is more unfamiliar is their inclination-dependence; and what we have termed {\sl reverse} resonances would be the normal corotation and Lindblad resonances for a bar with retrograde pattern speed (as one can see from the $i$-dependence).  Finally, there are what we have termed  {\sl epicyclic} and {\sl tumbling} resonances, which are strongest at high inclinations (near $\pi/2$).

In a more general setting (Sec.\ref{sec:matrix}), these resonance relations are written as $2\Omega_b = n_1\Omega_1 + n_2\Omega_2$, where $n_1,n_2$ are integers, $\Omega_1$ is the frequency of radial motions, and $\Omega_2$ is the mean longitudinal frequency, which generalise $\kappa$ and $\Omega$, respectively.  These number pairs are included in the last column of the table.

\begin{figure}
    \centering
        \includegraphics[width=0.5\textwidth]{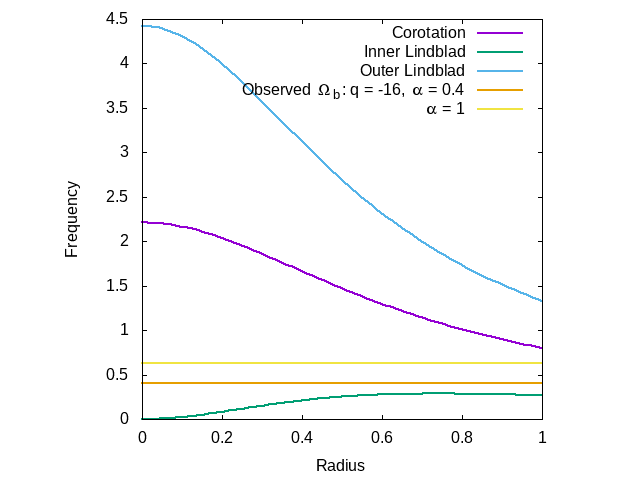}
    \caption{Resonances related to circular motions in a Plummer potential perturbed by a bar.   Also shown is the approximate pattern speed of a bar in a rotating model with $q = -16, \alpha = 1$ and 0.4.}
    \label{fig:resonances}
\end{figure}

The simplest situation in which to visualise the possible location of these resonances is the limiting case $q\to-\infty$ (the ``Einstein sphere"), when the frequencies $\Omega_1,\Omega_2$ become $\kappa,\Omega$. Combinations of these frequencies are plotted in Fig.\ref{fig:resonances}.  Also shown is the observed pattern speed (Sec.\ref{sec:temporalstructure}) found for the model with $q=-16$ in the cases $\alpha=0.4,1$.  Here the anisotropy is sufficiently tangentially biased that the Einstein sphere is approximately representative.

\begin{figure}
    \centering
        \includegraphics[width=0.5\textwidth]{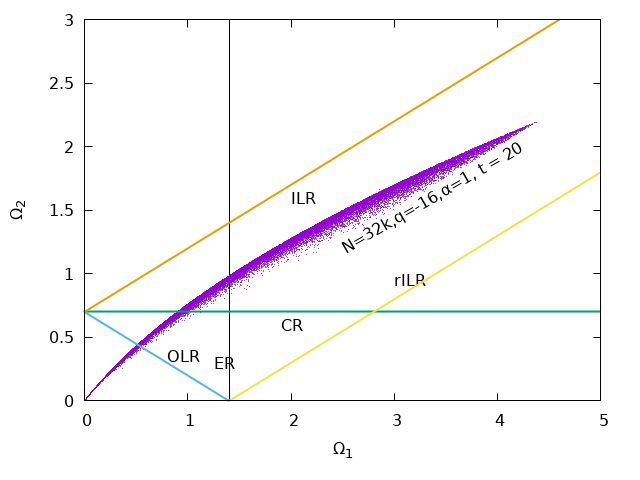}
    \caption{Distribution of  frequencies $\Omega_1,\Omega_2$  \dch{at time 15} in an $N$-body model with $N = 32768, q=-16,\alpha=1,\gamma_1=1$, along with the loci of five resonances.  \dch{At this time the model exhibits a growing} bar (Sec.\ref{sec:nbody-resonances}), and the pattern speed was determined in its early growth phase. OLR/ILR: Outer/Inner Lindblad Resonance, CR: Corotation Resonance, ER: epicyclic resonance (Table \ref{tab:resonances}), rILR: reverse ILR.}
    \label{fig:Lindblad}
\end{figure}

In a later section (Sec.\ref{sec:nbody-resonances}) we shall present some evidence of the influence of resonances on the behaviour of particles in $N$-body models.  Here, for one model, we simply present 
\sr
{in Fig.\ref{fig:Lindblad}} the distribution of particles in the space of the frequencies $\Omega_1,\Omega_2$, where their proximity to the various resonances can be appreciated visually. For the initial conditions of a model with $N=32768$ particles, $q=-16$ and $\alpha=1$ ($\gamma_1 = 1$ in Sec.\ref{sec:ics}), the values of $\Omega_{1,2}$ were calculated by integrating the equations of motion in the Plummer potential.

{}

The distribution is narrow because the anisotropy is quite extreme ($q=-16$) (in the limit $q\to-\infty$ the distribution would have been one-dimensional).  The Inner Lindblad Resonance and its retrograde counterpart appear to involve no particles, but we shall see (Sec.\ref{sec:application}) that the former still appears to play a central role in the dynamics, \dch{and present below a possible reason for this}.  There are, however, particles lying close to the other three resonances, and empirical evidence for their influence is studied in Sec.\ref{sec:nbody-resonances}.  
We can see immediately, however, that these particles are confined to the low-frequency part of the distribution, and this means that they affect stars at relatively large radius.  With the exception of the ``tumbling" resonance, the remaining resonances in Tab.\ref{tab:resonances} appear to play no role. 

\dch{In this section we identified resonances by examining the Fourier decomposition of the perturbing potential.  In first-order orbital perturbation theory, a given Fourier term gives rise to perturbations containing a resonant denominator $\omega - n_1\Omega_1-n_2\Omega_2$, where $\omega$ is the frequency of the explicit time-dependence of the perturbing potential.  If resonance is strictly defined as the vanishing of this denominator \citep[][p.810]{JalHun05} then resonances do not occur if the perturbation grows, but ``only near-resonances at which the denominators [....] are small" (if the growth-rate is small).  But these authors  also (their p.819) point to a situation, closely analogous to ours, in which the Fourier component corresponding to the ILR is important even though no particle is in strict or near-resonance.  It is a situation in which ``the denominator [...] is never large for any orbit", an apt description for the fact that the line labelled ``ILR" in Fig.\ref{fig:Lindblad} is approximately parallel to the distribution of points.  Such a Fourier term, it seems, may have a comparable importance to a resonant term in which a small fraction of particles are in ``near-resonance" but for most particles the denominator is large. Though it can be argued that there is no resonance at all if the denominator is not small compared with typical orbital frequencies, we prefer to regard the ILR as a weak resonance affecting most, if not all particles.  In general, in what follows we take a liberal view of what is meant by ``resonance".}

\subsection{Analysing the role of resonances with the matrix method}\label{sec:matrix}

Not all theoretical discussions of bar instabilities in disks seem readily adaptible to spherical systems, and here we follow specifically 
the  arguments  proposed  by \cite{Palmer+1989} and \cite{Allen+1992}.  These two papers were concerned largely with two instabilities.  In the earlier paper they developed a theory of what they termed the {\sl circular orbit instability}, generated by the inner Lindblad resonance, and in the later paper they turned to a so-called {\sl tumbling instability}, generated by what we have (for that reason) termed the tumbling resonance (Table \ref{tab:resonances}). 
Both papers proceed from the study of the linearised, orbit-averaged collisionless Boltzmann - Poisson system. More precisely, they demonstrate the existence of neutral (or unstable) solutions to the resulting eigensystem by only keeping 
terms that correspond to a particular resonance (or a particular set of resonances) between the instability and the orbital frequencies in the system. We propose to use a similar method, and try to test whether these mechanisms can be responsible for instabilities in the  systems studied in the present paper.

\subsubsection{Method}\label{sec:method}

In order to investigate in detail the role of resonances in the development of instabilities in tangentially anisotropic, rotating systems, we \dch{extended} the version of the matrix method formulated in \cite{Rozier+} \dch{to the case of} rotating spheres. The matrix method 
identifies unstable modes as eigenvectors (with an eigenvalue equal to 1) of the response matrix
\begin{equation}
    \widehat{M}_{pq} (\omega) \!=\! (2 \pi)^{3} \!\! \sum_{\mathbf n \in \mathbb{Z}^{3}\,} \!\! \int \!\! \mathrm d \mathbf J \, \frac{\mathbf n \cdot \partial F / \partial \mathbf J}{\omega - \mathbf n \cdot \mathbf{\Omega} (\mathbf J)} \big[ \psi^{(p)}_{\mathbf n} (\mathbf J) \big]^{*}\! \psi_{\mathbf n}^{(q)} (\mathbf J),
    \label{eq:RespMat}
\end{equation}
where the integral runs over the action space $(J_r, L, L_z)$ consisting of the radial action, the total angular momentum, and its $z$-component, respectively; $F$ is the phase-space distribution function of the system; $\mathbf \Omega = (\Omega_r, \Omega_{\phi}, 0)$ are the frequencies of a given orbit; $\psi^{(p)}_{\mathbf n}$ is a Fourier-transformed element of a bi-orthogonal potential-density basis; $\mathbf n$ is a three-vector of integers; and $\omega = \omega_0 + \mathrm i \eta$ is the (complex) frequency of the instability (oscillation frequency and growth rate). Examining the integrand in \dch{E}q.~(\ref{eq:RespMat}), we can see that the denominator is resonant, in the sense that it can take very small values in some parts of action space. Those orbits where $\omega - \mathbf n \cdot \mathbf{\Omega}(\mathbf J)$ is close to 0 are expected to be decisive in the development of the instability, since they contribute the most in the response matrix. We then call the 3-dimensional vector $\mathbf n$ of integers a \textit{resonance vector}: at a given $\mathbf n$, the corresponding integral will mainly involve the orbits that are at the $\mathbf n^{\text{th}}$ resonance with the instability. 

As a consequence, each integral term in the sum of \dch{E}q.~(\ref{eq:RespMat}) can be regarded as the contribution to the response matrix from the orbits that are at the $\mathbf n^{\text{th}}$ resonance with the instability. So, in order to investigate the role of each resonance in the development of instabilities, one can separately study the influence of each corresponding term in the matrix. We decided to follow this principle in order to study the interplay between resonances and instabilities in tangentially biased, rotating systems. More specifically, instead of just comparing the amplitude of the resonance terms in \dch{E}q.~(\ref{eq:RespMat}), we opted for the full mode search developed in \cite{Rozier+}, applied to response matrices in which only a subset of the resonance terms were kept: if $\mathcal I$ is a sub-ensemble of $\mathbb{Z}^{3}$, we searched for modes assuming the response matrix is restricted to
\begin{equation}
    \widehat{M}_{pq} (\omega) \!=\! (2 \pi)^{3} \!\! \sum_{\mathbf n \in \mathcal I\,} \!\! \int \!\! \mathrm d \mathbf J \, \frac{\mathbf n \cdot \partial F / \partial \mathbf J}{\omega - \mathbf n \cdot \mathbf{\Omega} (\mathbf J)} \big[ \psi^{(p)}_{\mathbf n} (\mathbf J) \big]^{*}\! \psi_{\mathbf n}^{(q)} (\mathbf J).
\label{eq:RespMattrunc}
\end{equation}
\dch{Such an approach has already been used in a study of disk instabilities by \citet{Poly2005}.}

An important constraint on the ensemble of resonances to consider comes from the fact that ${\widehat{M}_{pq} (\omega) \propto \delta_{m^p}^{n_3} \delta_{m^q}^{n_3}}$. Since we are focusing on 2-fold symmetric instabilities\footnote{\sr{Note that in principle, other instabilities could exist (i.e. with $m = 0$, giving axisymmetric oblate/prolate oscillations), however they are beyond the scope of this paper.}}, we only compute matrix terms with ${m^p = m^q = 2}$, hence the sum in \dch{E}q.~(\ref{eq:RespMattrunc}) is restricted to ${n_3 = 2}$. Consequently, in the following paragraphs, we will consider $\mathcal I$ as a sub-ensemble of $\mathbb{Z}^{2}$, and define ${\widetilde{\mathbf n} = (n_1,n_2)}$.

Using this notation, we can label some crucial resonances by their resonance vector ${\widetilde{\mathbf n}}$: (-1,2) corresponds to the inner Lindblad resonance, (0,2) is the corotation resonance, (1,2) is the outer Lindblad resonance, and so on (Table \ref{tab:resonances}). Each resonance should be understood through the relative value of the vector elements, so for each resonance vector ${\widetilde{\mathbf n}}$, its opposite ${-\widetilde{\mathbf n}}$ corresponds to the same kind of resonance\footnote{Note that the role of opposite resonance vectors is not the same; they only correspond to the same \textit{kind} of resonances. This is illustrated by the different inclination-dependence of resonances and their ``reverse" analogues in Table \ref{tab:resonances}.}.  Hence, whenever a resonance vector will be included in the ensemble $\mathcal I$, it will be assumed that its opposite is also included. As identified in \cite{Rozier+}, there is little influence from $|n_1| > 2$ to the instabilities in tangentially anisotropic, rotating systems, so that we will restrict the study to the pairs $\tbn$ with $|n_1| \leq 2$. As a consequence, the case of the complete matrix has eight different resonance vectors: $\mathcal{I}_0 = \{(0,0),(0,2),(-1,2),(1,0),(1,2),(-2,2),(2,0),(2,2)\}$, the first five of which are listed in Table \ref{tab:resonances}. 

Finally, we draw attention to the fact that  a major difference between rotating and  non-rotating systems emerges in the case ${\widetilde{\mathbf n} = (0,0)}$. 
The corresponding term in the response matrix (\alv{E}q.(\ref{eq:RespMat})) is directly related to the $L_z$-gradient of the distribution function, and is independent of the orbital frequencies. We will refer to it as the tumbling term: while it does not correspond to any resonance involving orbital frequencies, it is shown in \cite{Allen+1992} that this term is associated to the tumbling of orbital planes under the effect of a perturbation \citep[see also][]{Palmer1994}.

In summary, our aim is to investigate the role of resonances by examining the effect on the growth rate and pattern speed of removing some of the corresponding terms from the calculation.
In order to strengthen the case for this method, we present in Appendix~\ref{app:ROImatrix} the result of its application to the well-studied radial orbit instability regime found in \cite{Rozier+}.  We now apply it to the tangentially-biased rotating systems which are the focus of the present paper, and concentrate on the models with ${q=-16}$ and $\alpha = 0.4, 1$.

\subsubsection{Application to instability in tangentially-biased systems}\label{sec:application}

In order to identify which resonances are the most critical to the instabilities, we compared the results of the full matrix with results when a single resonance is removed. The importance of the removed resonance in the instability is then estimated through the change in growth rate and oscillation frequency of the instabilities found with and without the resonant term.

Table~\ref{tab:TANtruncrm} shows the values of the growth rates and pattern speeds that were obtained by the restricted matrix method in the two cases $\alpha = 0.4$ and $\alpha = 1$, alternatively removing each of the resonant terms that are present in the matrix of reference. Notice that the values of the growth rate and pattern speed of reference do not exactly correspond with those of \cite{Rozier+}: for the sake of numerical efficiency, we opted for fewer radial basis functions, so the mode reconstruction is not as accurate. However, the present reference case identifies the same instability, and we ensure the self consistency of the present method by performing matrix calculations with constant parameters, in particular the nature and number of radial basis functions. Similarly, the discrepancy in the measured growth rate of the $\alpha = 0.4$ cluster between simulations (see Fig.~\ref{fig:temporalstructure}) and matrix method is probably due to the relatively low number of basis functions used.

\begin{table}
	\centering
	\caption{Values of the growth rate and pattern speed found through the restricted response matrix method applied to the $(q, \alpha)=(-16, 0.4)$ and $(q, \alpha)=(-16, 1)$ tangentially-biased systems: comparison between the matrix of reference (complete) and the series of matrices obtained by removing a single resonance.}
	\begin{tabular}{ccccccc}
		\hline
		$\mathcal{I}_0 \setminus \mathcal{I}$ (removed) &  & $\eta(0.4)$ & $\Omega_b(0.4)$ &  & $\eta(1)$ & $\Omega_b(1)$ \\
		\hline
		$\varnothing$ (reference) &  & 0.012 & 0.36 &  & 0.24 & 0.66 \\
	    $(-1,2)$ &  & 0.0002 & 0.13 &  & 0.0066 & 0.33 \\
		$(1,2)$ &  & 0.0082 & 0.33 &  & 0.088 & 0.60 \\
    	$(0,2)$ &  & 0.0040 & 0.30 &  & 0.022 & 0.49 \\
		$(0,0)$ &  & 0.0073 & 0.33 &  & 0.044 & 0.40 \\
		$(1,0)$ &  & 0.0044 & 0.36 &  & 0.20 & 0.66 \\
		$(-2,2)$ &  & 0.010 & 0.36 &  & 0.24 & 0.66 \\
		$(2,2)$ &  & 0.010 & 0.33 &  & 0.20 & 0.66 \\
		$(2,0)$ &  & 0.011 & 0.36 &  & 0.24 & 0.66 \\
		\hline
	\end{tabular}
	\label{tab:TANtruncrm}
\end{table}

Let us first focus on the $\alpha = 0.4$ case. Some resonances have little influence both on the growth rate and on the pattern speed of the mode: removing either of $(-2,2)$, $(2,2)$ or $(2,0)$ does not impact the frequencies by more than 20\%\footnote{It is not hard to show that these terms are smaller than the others by a factor at least of order the epicyclic amplitude, which is small in these tangentially anisotropic models.}. Some resonances have a more significant impact on the instability growth rate: removing a term among $\{(1,2),(0,2),(0,0),(1,0)\}$ depletes the growth rate by at least 30\%, and up to 70\%. The pattern speed, however, is barely impacted by the removal of any of these terms. It appears that the contribution to the instability from the resonances in these groups can be interpreted as follows: none of them has a fundamental role in the formation of the instability, however they all contribute more or less to it by increasing its growth rate. This analysis does not apply to the ILR term $(-1,2)$: when it is removed, the remaining instability has a growth rate about 50 times smaller than the reference one, while its pattern speed is depleted by about a factor 3. This indicates that the instability identified in the absence of the ILR term is of a different nature from the one identified in all the other cases. 

The maximally rotating, $\alpha = 1$ cluster seems to present a more subtle behaviour. One can still identify a group of resonances that have little significance to the instability: $\{(1,0),(-2,2),(2,2),(2,0)\}$. The group of resonances that influence the growth rate but not the pattern speed is now reduced to one element: the removal of the OLR term results in the depletion of the growth rate by a factor 3 and of the pattern speed by about 10\%. A new category emerges: the tumbling and corotation terms are now associated to decreases of a factor respectively about 5 and 10 in the growth rate, and of respectively about 40\% and 25\% in the pattern speed. Finally, the ILR term is still the one bearing the most importance in the instability, as its removal leads to a depletion of a factor about 40 in the growth rate, and about 2 in the pattern speed. 

The most important process that leads to instability seems, in both $\alpha = 0.4$ and $\alpha = 1$ cases, to be borne by the ILR term in the matrix. This is indicative of a similarity between the instabilities in both cases. However, the behaviour of the other terms changes when rotation increases: the importance of the $(-2,2)$, $(2,2)$, $(2,0)$ and especially $(1,0)$ terms is lower at higher rotation, while corotation, the OLR and the tumbling terms gain importance with rotation. The tumbling term seems to gain a particular role in setting the high pattern speed of the instability at high $\alpha$.

Before we pass on from the matrix method, it is worth mentioning that we have focused here on the most rapidly growing mode.  For a broad ranges of values of large $\alpha$ and negative $q$ (in models where the Lynden-Bell demon is applied everywhere), it coexists with a second mode which has smaller growth rate  and pattern speed (by factors of order five and two, respectively).

\subsection{Resonances in an $N$-body model}\label{sec:nbody-resonances}

In this section we return to $N$-body modelling to examine evidence for resonant behaviour.  In particular we discuss direct $N$-body simulations of basic models with $\gamma_1=1$ (i.e. the Lynden-Bell demon is applied everywhere), especially the case of extreme tangential anisotropy ($q=-16$) and maximal rotation ($\alpha=1$).   This exhibits a rapidly growing bar which reaches maximum amplitude at about $t = 30$, but its growth is clearly nonlinear by $t = 15$  (when a plot of the amplitude against time has an inflexion from concave upwards to concave downwards).  During this shorter interval the pattern speed $\Omega_b$ decreases from 0.73 to 0.60.

\begin{figure}
    \centering
    \includegraphics[width=0.4\textwidth]{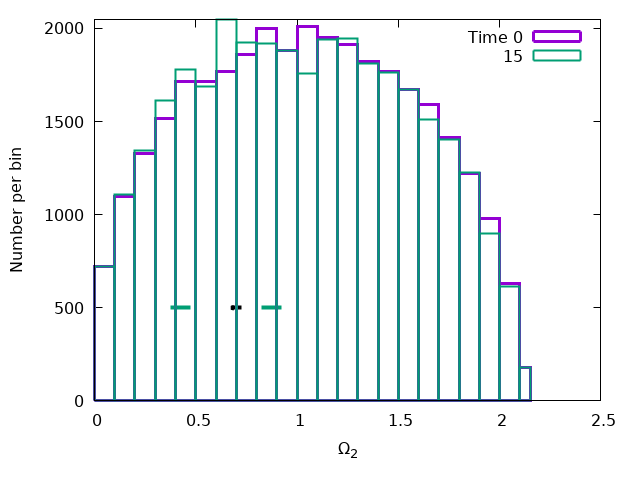}
    \caption{Evolution of the distribution of $\Omega_2$ during the early growth of a bar in an $N$-body model with $N=32768$ particles, $q=-16$ and $\alpha=1$.  The black bar  shows the variation of $\Omega_b$ (which decreases slightly) during the same interval of time.  The green bars show the ranges of the epicyclic (right) and Outer Lindblad (left) resonances.  For those resonances the breadth of frequencies is mainly due to the width of the distribution of frequencies in the initial conditions (Fig.\ref{fig:Lindblad}).}
    \label{fig:pf-cr}
\end{figure}

The initial conditions and relevant resonances are shown in Fig.\ref{fig:Lindblad}, and now Fig.\ref{fig:pf-cr} shows the evolution of the distribution of the azimuthal frequency $\Omega_2$ during the period of roughly linear behaviour.  In most cases it is difficult to distinguish features from sampling fluctuations, and we shall improve the statistics below, but in view of the large numbers of particles per bin, it is hard to dispute that there is a significant increase around $\Omega_2 = 0.65$ and an equally significant decrease around $1.05$ (the mid-point of the histogram bar is quoted). It is tempting to link the first of these with corotation. Again tentatively, the second could be associated with the ``epicyclic" resonance: in Fig.\ref{fig:Lindblad} the corresponding line crosses the upper margin of the distribution of points, where they are concentrated, at a frequency just below $\Omega_2 = 1$.  Other features at higher frequency, if regarded as significant, correspond to none of the resonances listed in Tab.\ref{tab:resonances}, but there is another at a lower frequency, about $\Omega_2 = 0.4$, which could very well correspond to the outer Lindblad resonance, as can be seen from Fig.\ref{fig:Lindblad}.  

\begin{figure}
    \centering
        \includegraphics[width=0.4\textwidth]{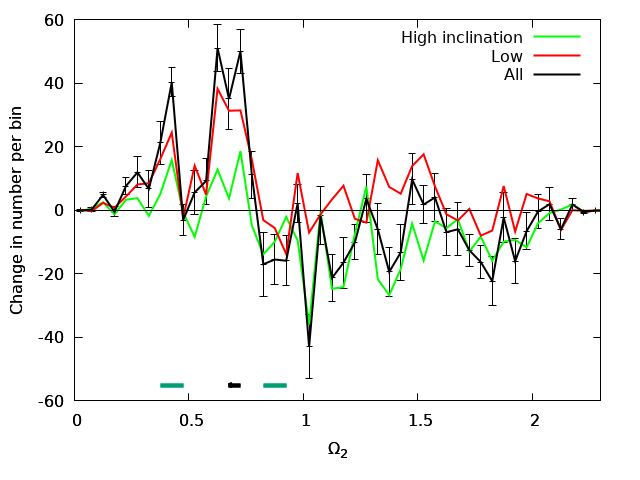}
    \caption{Change in the distribution of $\Omega_2$ between $t=0$ and $t=15$, given separately for particles of high ($i>60^\circ$; green) and low inclination (red); the thick black line gives the sum, i.e. all inclinations.  The error bars on the latter give the average and 1-$\sigma$ standard error of the mean over 10 simulations.  For the first two graphs the error bars would be larger by about 40\%.  The horizontal bars at the foot have the same meaning as in Fig.\ref{fig:pf-cr}, but depend on the bar pattern speed, which was determined for only one of the simulations. The binwidth is half that of Fig.\ref{fig:pf-cr}.  }
    \label{fig:inc}
\end{figure}

It is worth examining the distribution of those stars which contribute most to the evolution of  $f(\Omega_2)$.  Fig.\ref{fig:inc} shows two plots corresponding, respectively, to low and high inclination, and also the plot for all inclinations (black).  The red curve, for instance, shows the change from $t=0$ to $t=15$ in the number of particles in a bin (centred at the indicated values of $\Omega_2$), counting only those of inclination $i<60^\circ$.  Results from ten independent simulations have been combined, and the bin size is half that of Fig.\ref{fig:pf-cr}.  The sum, plotted in black, adds considerable significant detail to what  
can be inferred from Fig.\ref{fig:pf-cr}.  In the features at $\Omega_2\simeq 0.4$ and 0.65 the change is predominantly due to particles of low inclination, but the opposite is true of the feature at $\Omega_2\simeq1.05$. 
Bearing in mind the inclination-dependence noted in Tab.\ref{tab:resonances}, these results lend weight to the interpretations as behaviour at the outer Lindblad, corotation and epicyclic resonances, respectively.

Despite what has just been said about the epicyclic resonance, actually the resonant frequency is somewhat too small (see the position of the rightmost horizontal bar in the figures), and in this respect a better interpretation is offered by consideration of the ILR, the importance of which has already been established in Sec.\ref{sec:matrix}. It is defined by $\Omega_b = 
\OILR\equiv\Omega_2 - \Omega_1/2$, and it is found that, in the limit $q\to-\infty$, the expression $\OILR$ reaches its maximum value at a radius where $\Omega_2 = 1.09$.  Its importance is emphasised again in Sec.\ref{sec:mechanisms}.  A third  possible interpretation is that the feature near $\Omega_2 = 1$ (from about 0.8 to 1.1) is composite. 

As with the inclination, one may study the radial distribution in the bins which we have associated with resonance, but the radius is essentially given by the frequency $\Omega_2$, which is a decreasing function of radius.  Indeed the bins at $\Omega_2 = 0.4, 0.65$ and 1.05 correspond to approximate radial ranges $2<r<2.8, 1.2<r<1.6$ and $0.6<r<0.8$ respectively.

What has been learned from studying the evolution of the distribution of $\Omega_2$ can also be attempted with the distribution of $\Omega_1$, related to the epicyclic resonance, and $\Omega_1+2\Omega_2$, involving the OLR.  But again because of the narrowness of the distribution of the two basic frequencies, one sees evidence of rapid evolution at essentially the same corresponding frequencies, i.e. those which may be associated with all three resonances, plus some evidence for unattributable evolution at still larger frequencies.

\section{Discussion, Summary and Conclusions}

\subsection{Destabilising mechanisms in tangentially anisotropic, rotating spheres}\label{sec:mechanisms}

\subsubsection{A Theoretical Framework for Bar Formation}\label{sec:framework}

This paper has presented a diversity of empirical information on bar formation in a tangentially anisotropic, rotating spherical system.  To tie everything together, it is helpful to set our findings within a theoretical framework, and the one which we now present, though not original, has proved most fruitful in the exploratory aspects of our research.

\cite{Palmer+1989} and \cite{Allen+1992} exhibit two particular mechanisms that can give rise to linear instability in, respectively, tangentially anisotropic and rotating spherical stellar systems: the circular orbit instability and the tumbling instability.  Their approach is via a careful analysis of the linearised orbit-averaged Boltzmann-Poisson system, in which the first step is to treat one resonance in isolation, and then all others are treated as perturbations.  For both mechanisms, it is shown in the first step that the system would exhibit a particular set of neutral modes (i.e. modes with vanishing growth rate) {\sl in the absence of the perturbing resonances}, but that such a mode may be destabilised by the perturbing resonant interactions.   While the tumbling process is linked to the precession of orbital planes under the influence of a rotating bar, the circular orbit instability emerges from the influence of a bar on the apsidal precession of the orbits, entailing changes in $\OILR$ (defined in \alv{E}q.(\ref{eq:oilr})).

In the case of the circular orbit process, the characteristics of the instability are the following: 
\begin{itemize}
    \item The ILR is at the source of the neutral mode, hence it is the fundamental cause of the instability.
    \item The mode rotates at a pattern speed larger than (but close to) the maximum value of the ILR frequency, defined by \begin{equation}
        \OILR = \Omega - \frac{1}{2} \kappa.\label{eq:oilr}
    \end{equation} Fig.~\ref{fig:resonances} shows this frequency in the present clusters as a function of radius.
    \item Destabilisation is ensured by other resonances, at the oscillation frequency of the mode.
\end{itemize}
A similar list of characteristics can be given for the tumbling instability:
\begin{itemize}
    \item The tumbling process (corresponding to the $(0,0)$ resonance vector) is at the source of the neutral mode.
    \item The pattern speed is nearly proportional to the rotation parameter $\alpha$. This comes from the fact that the tumbling term in the eigenequation describing the neutral mode \citep[see \dch{e}q.~(8.31) in][]{Palmer1994} is proportional to $\alpha / \omega$.
    \item Destabilisation is ensured by other resonances, at the frequency of the mode.
\end{itemize}

Before passing on to a discussion of empirical evidence on the mechanisms of bar formation, we mention here one factor which supports the framework of the above theory: the presence of a second mode (see the last paragraph of Sec.\ref{sec:application}).  The main problem of the cited authors' approach, in which only one resonance is included in the first instance, exhibits not one but a sequence of neutral modes, and there is no reason to suppose that only  one of them can be destabilised by perturbing resonances.

\subsubsection{The role of tumbling and resonances: empirical evidence}\label{sec:empirical-evidence}

The series of experiments using the matrix method that we reported in Section~\ref{sec:matrix} helps in associating the instabilities we exhibited through $N$-body experiments (mainly in Sec.\ref{sec:results})  with the processes of tumbling and circular orbit instability.  We begin mainly with the matrix results, and then pass on to an interpretation of the $N$-body data.

The $(q, \alpha) = (-16, 0.4)$ case is particularly eloquent : it is obvious (from Table \ref{tab:TANtruncrm}) that the instability mostly depends on the ILR; the pattern speed of the mode lies just above the maximum of $\OILR$ (see Fig.\ref{fig:resonances}); and the role of other resonances seems to be to accumulate the growth rate. The $(q, \alpha) = (-16, 1)$ case seems to follow a similar trend, yet less tightly matching the characteristics of the circular orbit instability: the ILR term still bears most of the instability; the pattern speed is twice as high as the maximum of $\OILR$; and yet some other terms play an important role, in particular the one associated with the tumbling process. In contrast to the conclusions of \cite{Allen+1992}, it seems that the instabilities in the present tangentially anisotropic, rotating systems cannot be attributed to the tumbling instability alone. Perhaps the high values of the pattern speed and the growth rate may diminish the accuracy of their perturbative approach. The results of Table~\ref{tab:TANtruncrm} rather point to a mixing of the circular orbit and the tumbling processes, the ILR still being the most important resonance
\dch{in the creation of the instability}.

It is at first sight paradoxical to assert that the ILR is central to the observed instabilities, when Fig.\ref{fig:Lindblad} shows that \dch{no} particles in the models lie \dch{close to} 
resonance.   However, 
  the restricted matrix method (Table~\ref{tab:TANtruncrm}) shows that, even if the pattern speed is far away from any value $\OILR$ can take in the system (more than twice its maximum value \dch{when $\alpha = 1$}), the ILR is fundamental to the development of an instability in the system.  Furthermore there are indications in the $N$-body models that the ILR is playing a vital role:  the frequency measurements presented in Figs~\ref{fig:pf-cr} and~\ref{fig:inc} show strong activity in particles for which 
    $\Omega_2 = 1.05$, which corresponds closely to the value of $\Omega_2$  at which $\OILR$ takes its maximum value ($\OILR = 0.29$).  Finally on this point, we recall Fig.\ref{fig:temporalstructure}, which illustrates the ``floor" which the pattern speed reaches as the rotation parameter $\alpha$ decreases; this floor is near the maximum of $\OILR$, as expected in the scenario outlined in Sec.\ref{sec:framework}.

The foregoing remark, that a resonance may have a vital role even in the absence of resonant particles, \dch{echoes a point made near the end of Sec.\ref{sec:resonances-in-systems}}, and should be borne in mind in assessment of the role of the tumbling instability.  (This is formally the condition $\Omega_b = 0$, and is never satisfied in our models).  Consider in particular  
the region $\Omega_2 > 1.1$ in Figs~\ref{fig:pf-cr} and~\ref{fig:inc}, which corresponds to relatively low radii.  It is populated with orbits that do not resonate with the pattern at any of the usual low order resonances, 
but there is evidence in 
Fig.~\ref{fig:inc} that the variations in this inner region mostly affect high inclination orbits.  Therefore we interpret the activity here as being due to the tumbling process, because of its inclination-dependence (Table \ref{tab:resonances}). 

The results of Sections~\ref{sec:prograde-results} and~\ref{sec:high-shear-results} show that much importance attaches to these inner orbits: from the high-shear models, we know that flipping the orbits at $R<0.4$ is enough to nearly suppress the instability, while building a larger and larger rotating core of radius $0.4 < R < 0.75$ increases the growth rate of the instability in the inner system up to its maximum value. By contrast, the results of these two sections also show that what happens at larger radii ($R>0.5$, say) hardly matters:  this region may rotate, counter-rotate, or not rotate at all.  The importance of the most central orbits is also supported by the shapes of the modes, as shown by Fig.~\ref{fig:nbodyprojection} and in~\cite{Rozier+}: the modes seem to be confined to the inner part of the cluster, as by about $R=0.45$ in the present units, the density \dch{in the bar} has dropped to 10\% of its maximum value. 
All \dch{the} evidence supports the view that the tumbling process, while not dominant in the generation of instability, is an important component of the mechanism which creates it.

It should be noted that the foregoing discussion  refers to the cylindrical radius $R$. For high-inclination orbits, such as those most affected by the tumbling and epicyclic resonances, the spherical radius is significantly larger.
Still,
Fig.~\ref{fig:inc} shows that there is also some activity in the low inclination orbits (in the inner
region on which we have been focusing).

Low inclination orbits have a particularly  important role in the regions of the OLR and corotation  (around $\Omega_2 = 0.4$ and 0.7, respectively); these are orbits at relatively large radius.  The matrix results however show that the role of these resonances is of less importance than those of the tumbling or the ILR. This is consistent with the stability results of Section~\ref{sec:gamma34}: both high and low inclination orbits matter, but the high inclination ones are more important than the low inclination ones. 

A final factor which hints at the significance of the tumbling resonance at high $\alpha$ is the theoretical result (mentioned in Sec.\ref{sec:framework}) that the pattern speed of the relevant neutral mode is proportional to $\alpha$.  This suggests an explanation of the rising pattern speed for $\alpha\gtorder0.5$ in Fig.\ref{fig:temporalstructure}; this could not continue for small $\alpha$, as then the damping effect of the ILR would come into play \citep[][p.1292]{Palmer+1989}.  It is interesting to note also, from Table \ref{tab:TANtruncrm} in the case $\alpha=1$, that removal of the $(0,0)$ term reduces $\Omega_b$ to a value quite similar to the reference result for the case $\alpha = 0.4$ which, we have argued, is constrained by the maximum of $\OILR$.

To sum up, the present results point to the following scenario for the formation of instabilities in tangentially anisotropic, rotating systems: in the inner region of the cluster, the tumbling of orbital planes (especially for high-inclination orbits) and the drift in apsidal precession far from ILR cooperate to the formation of an unstable bar.
The growth rate of this pattern is enhanced by resonance with outer orbits at the epicyclic resonance and (most noticeably) the  corotation \dch{resonance} and the OLR, in particular with the low-inclination orbits. 
Thus there is a kind of cooperation between all the resonances, which diminishes high-inclination orbits in the inner regions, and enhances those of low inclination at larger radii.  Both effects create the flattened structure of the bar.

\subsection{Towards a general stability criterion}\label{sec:stability_criterion}

{In their classic study \citet{1973ApJ...186..467O} stated that disks become  ``approximately stable" when $T/\abs{W}$ falls to a value of $ 0.14 \pm 0.02$.  The wording shows just how difficult it is to establish stability through numerical methods, and the same problem arises in our work.  Furthermore, the emphasis in their work was on disks (with or without halos), whereas ours is mainly (but not exclusively) on spherical models which rotate globally.  Nevertheless we take the Ostriker \& Peebles criterion as our starting point in this subsection.}

\subsubsection{General rotating models studied in this paper}
 
Figs.~\ref{fig:opv1},~\ref{fig:opv1-gamma1} and~\ref{fig:opv.2v1-gamma1} summarise the results of our main numerical surveys, placing them 
in the space of growth rate, $T/W$ and particular circular velocities in the cluster. Even if there seems to be a trend towards more rapid instability with increasing $T/W$, an important fraction of our parameter space does not satisfy the Ostriker-Peebles criterion.
The present results simply add to the number of published deviations from this law \citep[e.g., see][]{ZanHoh78,BerMar79,Aoki79,Efs82,Evans98,Ath08}, and yet are among  the few that exhibit such deviations in purely spherical clusters. 

As a consequence, it appears that the linear stability of spherical systems cannot be uniquely determined by the ratio between global kinetic energy in rotating motion and global gravitational potential energy.   \dch{Nor would the global angular momentum be more useful:}   our simulations of clusters with counter-rotating components (high-shear models, Sec.\ref{sec:high-shear-results}) imply the presence of fast growing instabilities in clusters with arbitrarily low global angular momentum (when the angular momentum of the inner prograde and outer retrograde parts of the cluster compensate).

An important reservation should be made at this point, concerning the presence of instabilities with low growth rates. On one hand, in terms of measurements by analytical means (i.e. the response matrix in the present case), the behaviour of the analytical indicators at low growth rate gets very intricate \citep[see, e.g., ][]{PichonCannon1997,Merritt1999,Rozier+}, which makes it complicated to measure growth rates accurately. On the other, the measurement of slowly growing instabilities in numerical simulations also has a number of technical difficulties. In particular, measuring the growth of these very weak instabilities above the background noise requires a sufficient integration time,
and in $N$-body simulations as well as in actual physical systems, the time scale required for the growth of such instabilities can overlap with the secular time scale of  2-body relaxation in the cluster. As a consequence, one might miss the growth of an instability,
because the system has already secularly evolved to a stable configuration. 

Yet, even if we only focus on reasonably large growth rates (say $\eta > 0.05$), the results of Fig.~\ref{fig:opv1-gamma1} show that the stability of spherical clusters has at least one additional degree of freedom compared to the single one used by \cite{1973ApJ...186..467O}. As discussed in Section~\ref{sec:empirical-evidence}, rotation in the innermost part of the cluster seems to have a critical influence on the increase in growth rate, compared to the non-rotating case. Figure~\ref{fig:opv.2v1-gamma1} shows that a combination of $T/W$ and the mean azimuthal velocity at $R=0.2$ makes a better stability criterion than the original $T/W$, even if the separation between stable and unstable systems (or, rather, slowly and rapidly growing instabilities) in that space is not yet perfect. 

Let us remark in closing that the 2D space of global \dch{rotation} (through $T/W$) and azimuthal velocity at a given radius can be mapped into the space of global rotation and shear: a cluster with high $\vert\vphibar(0.2)\vert$ displays low shear if $T/W$ is also high, and high shear if $T/W$ is low; a cluster with low $\vert\vphibar(0.2)\vert$ displays high shear if $T/W$ is high, and low shear if $T/W$ is low.

\subsubsection{Non-rotating models with nearly circular motions}

\alv{We \dch{take the view} that the question of the stability of a non-rotating   
equilibrium with a high degree of tangential anisotropy is still unsettled. 
On the one hand, \cite{Pol87} uses the WKB approximation to show that such spheres are linearly stable. However, this result only probes the case of quasi-local disturbances, while we measure large scale patterns (as illustrated by Fig.\ref{fig:nbodyprojection}), which \dch{may be}  beyond the scope of the WKB approximation. On the other hand, the possibility of large scale instabilities in such spheres is supported by \cite{Palmer+1989}, based on complementary analytical and numerical results.}     A consequence of this finding, \dch{if confirmed}, would be that sufficiently tangentially anisotropic clusters seem to display unstable behaviour, \textit{whatever their degree of rotation or shear}. 
\sr{But  we \dch{caution}  that this intriguing regime deserves a careful analysis which goes beyond the scope of the current paper, and we hope to present a contribution dedicated to this topic in the future.}

\subsection{Conclusions}

This paper focuses on the collisionless stability of rotating, spherical stellar systems, using a mix of $N$-body simulations, the matrix method for the evolution of perturbations, and some physical ideas.  Our \alv{starting point} is a set of anisotropic Plummer models \citep{Dejonghe1987}, whose anisotropy is parametrised by a parameter $q$; our focus is on models with $q<0$, whose velocity distribution is tangentially biased.  To these models we apply different versions of the Lynden-Bell demon \citep{1960MNRAS.120..204L}, in which the rotation of a fraction $\alpha$ of stars is reversed in a region of phase space defined in terms of integrals of motion, and parametrised in various ways by a parameter $\gamma$.  Instances of these models were generated with the publicly available software \texttt{PlummerPlus}, and their evolution studied using the public version of \texttt{gyrfalcON}\footnote{https://teuben.github.io/nemo/man\_html/gyrfalcON.1.html}. Supplementary studies used \texttt{NBODY6} \citep{2012MNRAS.424..545N}. The matrix method, described for example in \citet{Rozier+}, was also used to search for unstable modes.

Our conclusions are as follows:
\begin{enumerate}
    \item For values of $q<0$ and $0<\alpha\le1$ the models are rather generally unstable to the formation of a rotating ``bar" perpendicular to the axis of rotation.
    \item In models in which the anisotropy parameter $q = -6$, and a fraction $\alpha$ of retrograde stars have the rotation reversed, the growth rate of the mode is approximately proportional to $\alpha^6$  (Fig.\ref{fig:m2d-v-op}), at least when $\alpha$ is large enough that the growth rate can be measured reliably.  In extremely tangentially anisotropic models, the pattern speed diminishes as $\alpha$ decreases, but remains above a certain limit, and is detectable even in non-rotating systems (Fig.\ref{fig:temporalstructure}).
    \item If only stars with energy below a certain fraction $\gamma_1$ of the minimum energy have their rotation reversed, the growth rate reaches roughly its maximum value if $\gamma_1\gtorder0.5$, i.e. if the Lynden-Bell demon is applied to at least the most bound half of the stars (Fig.\ref{fig:mumax}).
    \item In models in which stars below a given energy have the Lynden-Bell demon applied in a prograde sense and for the remainder it is applied in a retrograde sense, the minimum growth rate occurs when 
    the \dch{proportions} of retrograde and prograde stars in the most bound 40\% of the system \dch{are} nearly the same \sr
    { (i.e. the case $\gamma_2 = 0.2$ in Fig.\ref{fig:mumax1})}. 
    \item If the Lynden-Bell demon is applied to only the stars with orbital inclination above (or below) some limit, the growth rate continues to increase as the limit increases to include polar orbits (or decreases to include equatorial orbits, respectively) (Fig.\ref{fig:mumax34}).
    \item The bar produced by the instability is detectable only within the cylinder $R\ltorder1.5, \vert z\vert < 0.9$ in H\'enon units (Sec.\ref{sec:spatialstructure}).
    \item Though the dynamics is affected by several other resonances beyond the familiar Lindblad and corotation resonances, no individual resonance appears to play an essential role (Table \ref{tab:TANtruncrm}).
    \item The pattern speed of the bar, which varies surprisingly little from one model to another, is such as to avoid \sr
    { the inner Lindblad resonance everywhere, and} the corotation resonance in the cylindrical region where the bar is found (Figs.\ref{fig:temporalstructure},\ref{fig:resonances}).
    \item The inner Lindblad resonance is the single most important resonance in the bar instability, even though there are no particles at the ILR (Fig.\ref{fig:Lindblad} and Table \ref{tab:TANtruncrm}).  Other significant resonances are the outer Lindblad resonance, the corotation resonance, and two resonances which we refer to as ``tumbling" and ``epicyclic" resonances (Table \ref{tab:TANtruncrm} and Figs.\ref{fig:pf-cr} and \ref{fig:inc}).
    \item It cannot be shown, from the results in this paper, that any of the surveyed models is stable against bar formation.  If, instead of searching for a stability criterion, one searches for a criterion which determines which models may exhibit {\sl rapid} instability, we show that, in addition to the global rotational kinetic energy, it is helpful to include a measure of the rotation speed at a radius inside the virial radius of the system (Fig.\ref{fig:opv.2v1-gamma1}).
\end{enumerate}

\section*{Acknowledgements}

\alv{ We thank an anonymous \dch{r}eferee for \dch{a number of} helpful comments, including pertinent references.} SR thanks Christophe Pichon and Jean-Baptiste Fouvry for many interesting discussions, and all authors are grateful to them for commenting on an early draft. This work has made use of the Horizon Cluster hosted by Institut d'Astrophysique de Paris. All authors thank Stephane Rouberol for running smoothly this cluster.  We are grateful to S. Law and his IT staff in the Mathematics Department at Edinburgh University, for very helpful management of the GPU hardware.  Many simulations were also carried out with the ECDF cluster at the University.
All authors  thank the Leverhulme Trust for financial support from grant RPG-2015-408. ALV acknowledges support from a UKRI Future Leaders Fellowship (MR/S018859/1) and is grateful to the Department of Astronomy at the University of Tokyo, where part of this work was conducted, thanks to a JSPS International Fellowship with Grant-in-Aid (KAKENHI-18F18787).  

\subsection*{Data Availability}

Data underlying this article will be shared on reasonable request to the authors via d.c.heggie$@$ed.ac.uk. Data related to the initial conditions may be reproduced via the publicly available software {\tt  PlummerPlus}.








\appendix

\section{Determination of time scales in the models of Sec.\ref{sec:temporalstructure}}\label{sec:method-of-analysis}

This appendix describes details of the way in which data for Fig.\ref{fig:temporalstructure} were derived from results of $N$-body simulations.

The analysis of $C_2$
(\alv{E}q.(\ref{def_Cm})) began with the growth rate.  For large $\alpha$, $\abs{C_m}$ was found to saturate at some time $t_{max}$.  For these models ($\alpha\ge0.6$) the growth rate was obtained by fitting a logistic curve (which is nearly exponential at early times) in this time interval.  For smaller $\alpha$, where the growth of a bar (if present) did not saturate by time $t=200$, a simple exponential fit was used.  In these cases, however, data from several models were combined: eight for $\alpha = 0.4$ and ten each for $\alpha = 0,0.2$.  
In all cases, error bars represent the 1-$\sigma$ asymptotic confidence interval.

The pattern speed was estimated by examining the time series of $\arg(C_2)$.  For a rotating bar, this exhibits a sawtooth wave form, the angular speed of each segment being twice the pattern speed.  For large $\alpha$ it was evaluated by simply counting the number of periods within the selected time interval (see above), and the motion could be easily seen to be prograde.  For $\alpha=0.4$ the expected prograde pattern was still visible in places, but with irregularities which prevented the determination of the pattern speed by the same method.  For the pattern speed, data from several runs could not be simply combined, and an alternative method was used, viz. calculation of the autocorrelation of the time series of $\arg(C_2)$.  Averaging results over several runs (as mentioned above) gave a reliable autocorrelation with an estimate of the standard error.  The form of the average autocorrelation function is a clear decaying oscillation.  The location of its first maximum was determined by applying a quadratic fit to the section on either side of the maximum, as far as the neighbouring minima.  This method does not allow determination of the sense of pattern rotation.  This is therefore undetermined for $\alpha=0.2$, but is presumed to be prograde.  For $\alpha=0$ it may be either, or possibly a mixture (actually surface-harmonic expansion shows that there are five modes with $l=2$, but three are filtered out by the definition of $C_2$).

\section{Resonances in inclined epicyclic  motions perturbed by a rotating bar}\label{app:perturbation}

In the context of disk dynamics, 
\citet[][Sec.3.3.3]{2008gady.book.....B} use the expression $\Phi_1(R,\varphi) = \Phi_b(R)\cos[2(\varphi - \Omega_b t)]$ as a model for the potential of a planar bar,
where $R,\varphi$ are plane polar coordinates in the orbital plane, $\Omega_b$ is the pattern speed, and $\Phi_b$ specifies the radial dependence of the potential.  The obvious extension to three dimensions is $\Phi_1(r,\theta,\phi) = \Phi_b(r)\sin^2\theta\cos[2(\phi - \Omega_b t)]$ in spherical coordinates, since the angle-dependence is that of an $l = 2$ spherical harmonic.

We consider the first-order perturbing potential of a particle which, at lowest order, is in epicyclic motion on a plane inclined at inclination $i$ to the $x,y$ plane ($\theta=\pi/2$), and  passes its ascending node at longitude $\Omega$ at time $t = 0$\footnote{In the rest of the paper $\Omega$ means the circular angular velocity.}. In cylindrical polar coordinates $(R,\varphi,z)$ based on the orbital plane, with the angular coordinate $\varphi = 0$ at the node, it is easy (but a bit tedious) to show that
\begin{eqnarray}
    \Phi_1(R,\varphi,z=0) &=& \frac{1}{4}\Phi_b(R)\left\{
    2\sin^2i\cos[2(\Omega - \Omega_bt)] +\right.\nonumber\\
    &&(1+\cos i)^2\cos([2(\Omega-\Omega_bt+\varphi)]
    + \label{eq:potential}\\
    &&\left.(1-\cos i)^2\cos([2(\Omega-\Omega_bt-\varphi)]\right\}
    .\nonumber
\end{eqnarray}

Now we make use of the epicyclic approximation, by which \citep[][eqs.(3.91),(3.93a)]{2008gady.book.....B}
\begin{eqnarray}
   R &=& R_0 + X\cos\theta_1\label{eq:epicycle1}\\
   \varphi &=& \theta_2 - \gamma X\sin\theta_1,\label{eq:epicycle}
\end{eqnarray}
where $(R_0,\theta_2)$ are the cylindrical radius and longitude of the guiding centre, $X$ is the radial amplitude of the epicyclic motion, $\theta_1$ is the phase of the epicycle, and $\gamma$ is a certain function of $R_0$.  
Substituting \alv{E}qs.(\ref{eq:epicycle1}) and (\ref{eq:epicycle}) into \alv{E}q.(\ref{eq:potential}) and expanding to lowest order in $X$, we can readily locate all trigonometrical terms of the form 
\begin{equation}
    \left\{
    \begin{array}{c}
\cos\\
\sin
\end{array}
\right\}(2(\Omega-\Omega_b t) + n_1\theta_1 + n_2\theta_2),
\end{equation}
where $n_1,n_2$ are (small) integers. This includes the pair (0,0), which arises from the first term in \alv{E}q.(\ref{eq:potential}).  Using the fact that $\dot\theta_{1,2}=\Omega_{1,2}$ (the radial and azimuthal frequencies), these terms yield the resonances listed in Table \ref{tab:resonances}, including the stated $i$-dependence and the values in the last column.

\section{Application of the restricted matrix to the radial orbit instability}
\label{app:ROImatrix}

As described in \cite{2008gady.book.....B}, spherical systems biased towards radial orbits are subject to the so-called radial orbit instability. This instability emerges through a Jeans process, by the clustering of radially elongated orbits around a bar-like overdensity. More precisely, the torque applied by a weak bar on a radially elongated orbit acts to modify the apsidal precession rate of the orbit (i.e. $\Omega_2 - \Omega_1/2$), in such a way that the angle between the bar and the ellipse tends to decrease. This process results in a bar-like instability, with a pattern speed in the range of the radial orbits' precession rates. In this process, the orbits around the ILR are particularly critical, since they form the main body of the instability. In a rotating system, \cite{Palmer+1989} show that the tumbling process also participates in the destabilisation of the radial orbit instability. Here, we apply the restricted matrix method to the case of a radially anisotropic, rotating Plummer sphere, which gives a well constrained test-case for its application.

\begin{table}
	\centering
	\caption{Values of the growth rate and pattern speed found through the truncated response matrix method applied to the $q=2, \alpha=1$ radially-biased system.}
	\begin{tabular}{ccc}
		\hline
		$\mathcal I$ & $\eta$ & $\Omega_p$ \\
		\hline
		Reference  & 0.088 & 0.073\\
		$(-1,2),(0,0)$ & 0.088 & 0.066 \\
		$\mathcal{I}_0 \setminus (-1,2)$ & $<2 \times 10^{-4}$ & - \\
		$\mathcal{I}_0 \setminus (0,0)$ & 0.033 & 0.027 \\
		$(-1,2)$ & 0.018 & 0.024 \\
		$(0,0)$ & $<2 \times 10^{-4}$ & - \\
		\hline
	\end{tabular}
	\label{tab:ROItrunc}
\end{table}

Table~\ref{tab:ROItrunc} reports the values of the growth rate and oscillation frequency found by the restricted matrix method in the $q=2,\, \alpha=1$ system, when a variety of combinations of resonance vectors are tested. The computation of reference includes all relevant resonance vectors with ${|n_1|,|n_2| \leq 2}$, corresponding to ${\mathcal{I}_0 = \{(0,0),(1,0),(2,0),(-2,2),(-1,2),(0,2),(1,2),(2,2)\}}$.
It appears that turning off terms corresponding either to the ILR or to the tumbling process from the full matrix lower\dch{s} the growth rate and modif\dch{ies} the oscillation frequency of the instability, while keeping only those two terms together preserves both its growth rate and  (nearly) its oscillation frequency. These results show that the destabilisation of radially anisotropic rotating systems is guaranteed through the interaction of the radial orbit and tumbling processes only. In that case, the identification of these processes is made easy by the largely dominant role played by the corresponding terms in the response matrix: the second line of Table~\ref{tab:ROItrunc} shows that all other resonance vectors can be neglected w.r.t. $(-1,2),(0,0)$. Section~\ref{sec:matrix} shows that tangentially anisotropic systems are more complex in that respect.


\bsp	
\label{lastpage}
\end{document}